\documentclass[prd,preprint,
tightenlines,superscriptaddress,nofootinbib,showpacs]{revtex4}
\usepackage{amssymb,latexsym}
\usepackage{amsmath,amsbsy,bbm}
\usepackage{epsfig,bm}
\usepackage{graphicx,comment}
\usepackage{color}
\DeclareMathOperator{\tr}{tr}

\begin{document}
\def\a{{\alpha}}
\def\b{{\beta}}
\def\d{{\delta}}
\def\D{{\Delta}}
\def\e{{\varepsilon}}
\def\g{{\gamma}}
\def\G{{\Gamma}}
\def\k{{\kappa}}
\def\l{{\lambda}}
\def\L{{\Lambda}}
\def\m{{\mu}}
\def\n{{\nu}}
\def\o{{\omega}}
\def\O{{\Omega}}
\def\S{{\Sigma}}
\def\s{{\sigma}}
\def\th{{\theta}}
\def\X{{\Xi}}

\newcommand{\mnod}{\stackrel{\circ}{M}}

\def\ol#1{{\overline{#1}}}

\def\Dslash{\ol D\hskip-0.65em /}
\def\Dslashe{D\hskip-0.65em /}
\def\Pslash{\ol P\hskip-0.65em /}
\def\lslash{l\hskip-0.35em /}
\def\Pslashe{P\hskip-0.65em /}

\def\Dtslash{\tilde{D} \hskip-0.65em /}

\def\CPT{{$\chi$PT}}
\def\QCPT{{Q$\chi$PT}}
\def\PQCPT{{PQ$\chi$PT}}
\def\tr{\text{tr}}
\def\str{\text{str}}
\def\diag{\text{diag}}
\def\order{{\mathcal O}}

\def\cG{{\mathcal G}}
\def\cF{{\mathcal F}}
\def\cC{{\mathcal C}}
\def\cB{{\mathcal B}}
\def\cT{{\mathcal T}}
\def\cQ{{\mathcal Q}}
\def\cL{{\mathcal L}}
\def\cO{{\mathcal O}}
\def\cA{{\mathcal A}}
\def\cQ{{\mathcal Q}}
\def\cR{{\mathcal R}}
\def\cH{{\mathcal H}}
\def\cW{{\mathcal W}}
\def\cM{{\mathcal M}}
\def\cD{{\mathcal D}}
\def\cN{{\mathcal N}}
\def\cP{{\mathcal P}}
\def\cK{{\mathcal K}}
\def\Qt{{\tilde{Q}}}
\def\Dt{{\tilde{D}}}
\def\St{{\tilde{\Sigma}}}
\def\cBt{{\tilde{\mathcal{B}}}}
\def\cDt{{\tilde{\mathcal{D}}}}
\def\cTt{{\tilde{\mathcal{T}}}}
\def\cMt{{\tilde{\mathcal{M}}}}
\def\At{{\tilde{A}}}
\def\cNt{{\tilde{\mathcal{N}}}}
\def\cOt{{\tilde{\mathcal{O}}}}
\def\cPt{{\tilde{\mathcal{P}}}}
\def\cI{{\mathcal{I}}}
\def\cJ{{\mathcal{J}}}
\def\cb{{\cal B}}
\def\cbb{{\overline{\cal B}}}
\def\ct{{\cal T}}
\def\ctt{{\overline{\cal T}}}

\def\eqref#1{{(\ref{#1})}}

\preprint{MIT-CTP-4093}
\preprint{UMD-40762-458}
 
\title{Hyperon Electromagnetic Properties in Two-Flavor Chiral Perturbation Theory }

\author{Fu-Jiun~Jiang}
\email[]{fjjiang@ctp.mit.edu}
\affiliation{Center for Theoretical Physics, Department of Physics, Massachusetts Institute of Technology, Cambridge, MA 02139, USA}

\author{Brian~C.~Tiburzi}
\email[]{bctiburz@umd.edu}
\affiliation{Maryland Center for Fundamental Physics, Department of Physics, University of Maryland, College Park, MD 20742-4111, USA}

\date{\today}

\pacs{12.39.Fe, 14.20.Jn}

\begin{abstract}
The pion mass dependence of hyperon electromagnetic properties 
is determined using two-flavor heavy baryon chiral perturbation theory. 
Specifically we compute chiral corrections to the charge radii, 
magnetic moments, 
and magnetic radii of the spin one-half hyperons, 
as well as the 
charge radii, 
magnetic moments, 
magnetic radii, 
electric quadrupole moments, 
and quadrupole radii of the spin three-half hyperons. 
Results for the nucleon and delta are also included. 
Efficacy of the two-flavor theory is investigated by analyzing the role played by virtual kaons. 
For the electromagnetic properties of spin one-half hyperons, 
kaon loop contributions are shown to be well described by terms analytic in the pion mass squared. 
Similarly kaon contributions to the magnetic moments of spin three-half hyperons 
are well described in the two-flavor theory. 
The remaining electromagnetic properties of spin three-half resonances can be described 
in two-flavor chiral perturbation theory, 
however, 
this description fails just beyond the physical pion mass.
For the case of experimentally known hyperon magnetic moments and charge radii,
we demonstrate that chiral corrections are under reasonable control,
in contrast to the behavior of these observables in the three-flavor chiral expansion. 
The formulae we derive are ideal for performing the pion mass extrapolation
of lattice QCD data obtained at the physical strange quark mass. 
 \end{abstract}

\maketitle

\section{Introduction}
\label{introduction}

Before the advent of QCD, 
the spectrum of the lowest-lying hadrons provided a clue to the underlying flavor symmetries of the theory of strong interactions. 
The lowest-lying mesons and baryons seem to be organized into 
$SU(3)$ 
multiplets; 
and, 
the lightest states, 
the pseudoscalar mesons,  
are suggestive of spontaneously broken chiral symmetry, 
$SU(3)_L \times SU(3)_R \to SU(3)_V$. 
Explicit chiral symmetry breaking introduced by three light quark masses would give rise to small masses for the octet of pseudoscalar Goldstone bosons. 
From a modern perspective, 
such features of low-energy QCD can be explained using a model-independent framework that describes the interactions 
of the lowest-lying hadrons with the pseudo-Goldstone modes. 
This theory is chiral perturbation theory (\CPT)~\cite{Gasser:1983yg,Gasser:1984gg}; 
and, 
for three light quark flavors,
the lowest-lying baryons are grouped into multiplets of the unbroken $SU(3)_V$ symmetry. 
The
\CPT\ framework provides a rigorous description of low-energy QCD
provided the light quark masses, 
$m_u$, 
$m_d$, 
and 
$m_s$,  
are much smaller than the QCD scale, 
$m_u, m_d, m_s \ll \L_{QCD}$.

In the presence of electromagnetic interactions, 
$SU(3)$ 
restricts possible
baryon magnetic moment operators, 
for example.
Consequently relations between magnetic moments emerge. 
With vanishing quark masses, 
group theory permits only two magnetic moment operators for the octet baryons. 
Including the magnetic transition between the 
$\Sigma^0$ 
and 
$\Lambda$
baryons, 
there are eight known magnetic moments, 
and hence 
six 
$SU(3)$ 
symmetry relations are predicted~\cite{Coleman:1961jn}. 
In heavy baryon 
\CPT\ (HB\CPT)~\cite{Jenkins:1990jv,Jenkins:1991es},
these Coleman-Glashow relations emerge from the leading-order (LO)
operators in the chiral expansion.   
In Table~\ref{t:CGO},
we summarize the Coleman-Glashow relations. 
Experimentally the relations are reasonably well satisfied, 
and in some cases suggest that 
$SU(3)$ 
breaking could be treated perturbatively.

Baryon electromagnetic properties can be determined using 
$SU(3)$
HB\CPT\ 
beyond LO~\cite{Jenkins:1992pi,Bernard:1992xi,Meissner:1997hn,Kubis:1999xb}.
At next-to-leading order (NLO), 
the magnetic moments receive contributions that are non-analytic in the quark masses, 
without additional low-energy constants.%
\footnote{
Technically when the decuplet resonances are included, 
new NLO operators are possible. 
These new operators are merely the LO operators multiplied by the chiral singlet quantity 
$\Delta / \Lambda_\chi$, 
where 
$\D$ 
is the average splitting between the decuplet and octet baryons, 
and 
$\L_\chi$ 
is the chiral symmetry breaking scale. 
As the determination of such low-energy constants requires the ability to vary 
$\D$, 
we shall subsume such NLO operators into the LO ones. 
}
In this scheme, 
deviations from the Coleman-Glashow relations
first arise from the leading meson loop contributions, 
which depend on a few known axial coupling constants. 
These predictions are shown in Table~\ref{t:CGO}, 
with the relevant formulae collected in the Appendix. 
The agreement with experiment is not very good.
Large corrections at NNLO, which scale as 
the ratio of the kaon mass to octet baryon mass, 
$m_K / M_B \sim 0.45$, 
are certainly possible~\cite{Puglia:1999th}; 
however, 
one must go to NNNLO to determine if the expansion is truly under control. 
Nonetheless, 
if 
$SU(3)$ 
HB\CPT\
is converging for the Coleman-Glashow relations, 
it does so slowly. 
It could be possible that quantities protected from large 
$SU(3)$
breaking converge more quickly. 
At NLO, 
there are three relations between magnetic moments that are insensitive to the non-analytic
quark mass dependence~\cite{Caldi:1974ta,Jenkins:1992pi}. 
At NNLO, 
there are even two relations that eliminate the linear dependence 
on the strange quark mass~\cite{Okubo:1963zza}.
These higher-order relations are all well-satisfied experimentally. 
Unfortunately HB\CPT\ does not make parameter-free
predictions for these higher-order relations.

%
  \begin{table}[t]
  \begin{center}
     \caption{Relations between octet baryon magnetic moments. 
      The $SU(3)$ HB\CPT\ results quoted are NLO values for the numerator divided by experimental values for the denominator.  
     NLO expressions for the numerators are provided in the Appendix. 
     The $\D \%$ is the relative percent difference of the HB\CPT\ calculation compared to the 
     experimental value.}
   \smallskip
   \begin{tabular}{||c|c|c|c||}
    Relation & 
    $\quad$ Experiment  $\quad$ & 
    $\quad$ HB\CPT $\quad$ & 
    $\quad |\D \%| \quad$  \\
    \hline
    \hline
    LO (Coleman-Glashow) & 
    &
    &
    \\
    $ (\mu_{\S^-} -  \mu_{\X^-}  ) / ( \mu_{\S^-} + \mu_{\X^-}   ) = 0$ &
    $0.28$ & 
    $0.50$  &
    $77 \%$ \\
     $( \mu_n - 2 \mu_{\L}  ) / ( \mu_n  +  2 \mu_{\L} )= 0$ &
    $0.22$ & 
    $0.61$ &
    $180 \%$ \\
    $ ( \mu_n - \mu_{\X^0} )  /  ( \mu_n + \mu_{\X^0} ) = 0$ &
    $0.21$ &
    $0.58$ &
    $180 \%$ \\
     $ (\mu_p - \mu_{\S^+} ) / (\mu_p + \mu_{\S^+}) = 0$ &
    $0.064$ &
    $0.21$ &
    $230 \%$ \\
    $ ( \mu_{n} + \mu_{\Sigma^-} + \mu_{p}  ) / (  \mu_{n} + \mu_{\Sigma^-} - \mu_{p} ) = 0$ &
    $0.048$ &
    $0.14$ &
    $190 \%$ \\
    $ (\sqrt{3} \mu_{n}  +  2 \mu_{\Sigma\Lambda} ) /  ( \sqrt{3} \mu_{n}  -  2 \mu_{\Sigma\Lambda} ) = 0$ &
    $0.014$ &
    $0.15$ &
    $970 \%$ \\
    \hline
    NLO (Caldi-Pagels) &
    &
    &
    \\
    \large
    $ \frac{\mu_p + \mu_{\X^0} + \mu_{\X^-} + \mu_n - 2 \mu_\L}{\mu_p - \mu_{\X^0} - \mu_{\X^-} - \mu_n + 2 \mu_\L}$ \normalsize $=0$ &
    $0.038$ &
    $0$ &
    -- \\
    \large $\frac{\sqrt{3} \m_{\S \L}  + \mu_{\X^0} + \mu_n - \mu_\L}{\sqrt{3} \m_{\S \L}  - \mu_{\X^0} - \mu_n - \mu_\L}$ \normalsize $= 0$ &
    $0.036$ &
    $0$ &
    -- \\
    $( \mu_{\S^+} + \mu_{\S^-} + 2 \mu_\L) / ( \mu_{\S^+} - \mu_{\S^-} - 2 \mu_\L) = 0$ &
    $0.015$ &
    $0$ &
    -- \\
    \hline
    NNLO (Okubo) &
    &
    &
    \\
       $(\mu_{\Sigma^+} + \mu_{\Sigma^-} - 2 \mu_{\Sigma^0} ) / (\mu_{\Sigma^+} - \mu_{\Sigma^-} + 2 \mu_{\Sigma^0} ) $ = 0 & 
    -- &
    $0$  &
    -- \\
    \large
      $\frac{6 \mu_{\L} + \mu_{\Sigma^+} + \mu_{\Sigma^-} - 4 \mu_{\X^0} - 4 \mu_n - 4 \sqrt{3} \mu_{\S \L}}
    {6 \mu_{\L} - \mu_{\Sigma^+} + \mu_{\Sigma^-}  + 4 \mu_{\X^0} + 4 \mu_n - 4 \sqrt{3} \mu_{\S \L}}$
    \normalsize $= 0$ & 
    $0.028$ &
    $0$ &
    -- \\
    \hline
    \hline
       \end{tabular}
  \label{t:CGO}
  \end{center}
 \end{table}
%

Whatever the status of the Coleman-Glashow and higher-order relations between magnetic moments in \CPT, 
the 
$SU(3)$ 
chiral corrections to individual baryon magnetic moments do not appear to be under perturbative control.
Renormalizing the NLO loop contributions such that they vanish in the chiral limit
renders them scale-independent.
Hence we can compare just the loop contributions to the experimental 
moments to determine whether the perturbative expansion is under control. 
Defining the relative difference 
$\d \mu_B 
=  
| \mu^{loop}_B  /   \mu_{B} | 
$, 
we have tabulated the size of 
$SU(3)$ 
loop contributions to baryon magnetic moments in Table~\ref{t:loops}. 
The size of these loop contributions suggests that convergence of the 
$SU(3)$ 
chiral expansion for baryon magnetic moments is slow at best.


%
  \begin{table}[b]
  \begin{center}
     \caption{
     Relative size of NLO loop contributions compared to experiment in 
     $SU(2)$ and 
     $SU(3)$ HB\CPT. 
     }
   \smallskip
   \begin{tabular}{c|cccccccc}
    Theory &
    $\quad \d \mu_p \quad $ &
    $\quad \d \mu_n \quad $ &
    $\quad \d \mu_{\S^+} \quad $ &
    $\quad \d \mu_{\S^-} \quad $ &
    $\quad \d \mu_{\L} \quad $ &
    $\quad \d \mu_{\S \L} \quad $ &
    $\quad \d \mu_{\X^0} \quad $ &
    $\quad \d \mu_{\X^-} \quad $ \\
    \hline
    \hline
    $SU(3)$ HB\CPT &
    $66 \%$ &
    $41 \%$ &
    $120 \%$ &
    $21 \%$ &
    $220 \%$ &
    $74 \%$ &
    $210 \%$ &
    $176 \%$  \\
    $SU(2)$ HB\CPT &
    $39 \%$ &
    $57 \%$ &
    $17 \%$ &
    $35 \%$ &
    $0$ &
    $18 \%$ &
    $< 1\%$ &
    $<1 \%$ \\ 
    \hline
    \hline
       \end{tabular}
  \label{t:loops}
  \end{center}
 \end{table}
%

Large loop contributions to baryon magnetic moments can arise from virtual kaons, 
and attempts have been made to improve the 
$SU(3)$ 
expansion of baryon observables.
One approach is to use a long-distance regularization scheme
that subtracts different short-distance effects compared to dimensional regularization~\cite{Donoghue:1998bs}. 
Another approach is to treat the baryons relativistically. 
This is equivalent to the conventional heavy baryon approach with a resummation of a certain subset of diagrams. 
Results for baryon electromagnetic properties using the infrared regularization scheme of Ref.~\cite{Becher:1999he} 
show some improvement~\cite{Kubis:2000aa}. 
Recent work employing a different scheme, 
however, 
shows promising results for both octet and decuplet baryons~\cite{Geng:2008mf,Geng:2009hh,Geng:2009ys}.

Our approach to the problem is altogether different. 
We begin with the observation that it is possible to reorganize the
three-flavor chiral expansion into a two-flavor one thereby excluding the kaon and eta loops%
~\cite{Roessl:1999iu,Frink:2002ht,Beane:2003yx,Tiburzi:2008bk,Flynn:2008tg,Bijnens:2009yr,Jiang:2009sf,Mai:2009ce}. 
The 
$SU(2)$ 
theory  of hyperons exploits the hierarchy of scales
$m_u, m_d \ll m_s \sim \L_{QCD}$. 
Consequently the strange quark mass dependence  
is either absorbed into the leading low-energy constants of 
$SU(2)$, 
or arises through power-law suppressed terms, 
$\sim ( m / m_s)^n$,
which are absorbed into low-energy constants of pion-mass dependent operators.  
Here 
$m$ 
is used to denote the average of the up and down quark masses. 
The resulting theory sums all potentially large strange quark mass contributions to all orders. 
Improved convergence over 
$SU(3)$ 
has been explicitly shown for hyperon masses~%
\cite{Tiburzi:2008bk}, 
isovector axial charges~%
\cite{Jiang:2009sf}, 
and pion-hyperon scattering lengths~%
\cite{Mai:2009ce}. 
We undertake the study of hyperon electromagnetic properties 
in order to arrive at chiral corrections that are under better control perturbatively. 
The size of  
$SU(2)$ 
chiral corrections to baryon magnetic moments derived in this work are also shown in Table~\ref{t:loops}.
For hyperons, 
the results seem to indicate improvement over 
$SU(3)$ 
HB\CPT. 
Additionally expressions we derive for the pion mass dependence of
hyperon electromagnetic properties are ideal for performing lattice QCD extrapolations. 
In extrapolating lattice QCD data on hyperon properties, 
typically only the pion mass extrapolation is required as the strange 
quark mass is fixed at or near its physical value. 
Such extrapolations for strange hadrons are most economically done with 
$SU(2)$ 
\CPT.

For states lying above the nucleon isodoublet, 
efficacy of the two-flavor theory strongly depends on the underlying 
$SU(3)$ 
dynamics.%
\footnote{
This is true even for the quartet of 
delta-resonances.
The isosinglet
$\L$ baryon lies 
$0.12 \, \texttt{GeV}$ 
\emph{below} the 
delta 
multiplet. 
The non-analytic contributions to 
delta 
properties from 
$K \L$ 
intermediate states, 
however, 
are well described
by terms analytic in the pion mass squared,
and consequently non-analytic in the strange quark mass. 
}
Kinematically, 
hyperons are forbidden to produce kaons through strong decays.
The nearness of strangeness-changing thresholds, 
however, 
can lead to significant non-analytic quark mass dependence
in hyperon observables.
Such dependence may not be adequately captured in the two-flavor theory 
because explicit kaons are absent. 
Due to the size of hyperon mass splittings, 
spin three-half hyperon resonances are particularly sensitive to kaon contributions.
The 
$SU(2)$
chiral expansion of kaon loop contributions has been demonstrated 
to be under control for hyperon masses and isovector axial charges~%
\cite{Tiburzi:2009ab,Jiang:2009fa}.
Here we additionally explore the effects of virtual kaons on the 
electromagnetic properties of hyperons. 
For the spin one-half hyperon electromagnetic properties, 
kaon loop contributions are well captured by terms 
analytic in the pion mass squared. 
The same remains true for magnetic moments of the 
spin three-half hyperons. 
Electromagnetic radii and quadrupole moments of the hyperon resonances are shown to be
quite sensitive to the nearby kaon thresholds. 
The 
$SU(2)$ 
expansion of these kaon contributions appears to converge at the physical pion mass, 
however, 
the efficacy of the two-flavor theory does not extend considerably 
far beyond the physical point.

Our presentation has the following organization. 
In Sec.~\ref{su2L}, 
we review two-flavor 
HB$\chi$PT for hyperons and introduce electromagnetism into the theory. 
Using this two-flavor theory, 
we calculate the electromagnetic properties of the spin one-half and spin three-half hyperons in Sec.~\ref{spin-1/2}. 
We work at NLO in the chiral and heavy baryon expansions. 
Following that, 
we investigate the effect of virtual kaons on the various electromagnetic properties of hyperons in Sec.~\ref{discussion}. 
Additionally the size of chiral corrections to these observables is determined
by making contact with experimental data. 
The predicted pion mass dependence is compared with available lattice data.
Expressions from 
$SU(3)$  
HB\CPT\ needed for the remarks in the introduction have been collected in an Appendix. 
Finally Sec.~\ref{conclusion} concludes our study.

\section{$SU(2)$ HB\CPT\ for Hyperons}
\label{su2L}

In this section, 
we briefly review the 
$SU(2)$ 
effective Lagrangians for hyperons,
and importantly include electromagnetic interactions. 
We largely follow the formulation used in~\cite{Tiburzi:2008bk}.
For comparison purposes, 
we give first the well-known chiral Lagrangian for nucleons and deltas with electromagnetism. 
For ease, 
we write the local electromagnetic operators in a way which is general to each isospin multiplet.

\subsection{Strangeness $S=0$ Baryons \label{S=0L}}

At leading order, 
the effective Lagrangian for the nucleon and delta resonances including interaction terms with pions is given by~%
\cite{Bernard:1995dp,Hemmert:1997ye}   
\begin{eqnarray}
\label{eq:LS=0}
\cL^{(S=0)} 
&=& 
i \ol N v \cdot \cD N
- 
i \ol T_\mu v \cdot \cD \, T^\mu
+ 
\D \ol T_\mu T^\mu +
2 g_A 
\ol N S \cdot A N \nonumber \\
&+&
g_{\D N}
\left( \ol T_\mu A^\mu N + \ol N A_\mu T^\mu \right)+ 
2 g_{\D\D}
\ol T_\mu S \cdot A T^\mu,
\end{eqnarray}
where 
$N$ 
is the nucleon doubtlet 
$N = (p,n)^{T}$ ,
and the decuplet field 
$T_{ijk}$ 
is symmetric under any permutation of the indices 
$i,j,k \in \{1,2\}$.
The 
delta 
resonances are embeded in 
$T_{ijk}$ as 
$T_{111} =\Delta^{++}$, 
$T_{112} = \Delta^{+}/\sqrt{3}$, 
$T_{122} = \Delta^{0}/\sqrt{3}$, 
and 
$T_{222} = \Delta^{-}$. 
Further, 
the tensor products between nucleon and resonances are given by
$( \ol T  A \, T )
= 
\ol T {}^{kji}  A_{i} {}^{l} \, T_{ljk}$,
and 
$( \ol T A \, N )
=
\ol T {}^{kji}  A_{i} {}^{l} \, N_{j} \, \epsilon_{kl}
$.
The derivatives 
${\cal D}_{\mu}$ 
appearing in eq.~(\ref{eq:LS=0}) 
are both chirally covariant and electromagnetically gauge covariant, 
and act on
$N$ 
and 
$T$ 
fields in the following manner
\begin{eqnarray}
(\cD_\mu N)_i 
&=&  
\partial_\mu N_i  +  (V_\mu)_i {}^j  N_j + \tr ( V_\mu ) N_i ,   
\nonumber \\
\left( \cD_\mu T_\nu \right)_{ijk}
&=&
\partial_\mu (T_\nu)_{ijk}
+ 
(V_{\mu})_{i} {}^l  (T_\nu)_{lkj}
+
(V_\mu)_{j} {}^l  (T_\nu)_{ilk}
+
(V_\mu)_{k} {}^l  (T_\nu)_{ijl}
.\end{eqnarray}
The vector and axial fields of pions, 
namely 
$V_{\mu}$ 
and 
$A_{\mu}$,
in the presence of an electromagnetic gauge field,
$\cA_\mu$, 
are given by 
\begin{eqnarray}
V_\mu 
&=& 
\frac{1}{2} 
\left( \xi  \partial _\mu \xi^\dagger + \xi^\dagger  \partial_\mu \xi \right)
+ 
\frac{1}{2} i e \cA_\mu
\left( \xi  Q \xi^\dagger + \xi^\dagger Q \xi \right),
\label{eq:Vector}
\\
A_\mu
&=& 
\frac{i}{2} 
\left(  \xi \partial _\mu \xi^\dagger -  \xi^\dagger \partial_\mu \xi \right)
-
\frac{1}{2} e \cA_\mu
\left( \xi  Q \xi^\dagger - \xi^\dagger Q \xi \right)
\label{eq:Axial}
,\end{eqnarray}
with 
$\xi = \exp ( i \phi / f)$, 
where the pion fields 
$\phi$
live in an $SU(2)$ matrix 
\begin{equation}
\phi = 
\begin{pmatrix}
\frac{1}{\sqrt{2}} \pi^0  & \pi^+ \\
\pi^- & - \frac{1}{\sqrt{2}} \pi^0
\end{pmatrix}
,\end{equation}
and 
$f  = 130 \, \texttt{MeV}$ 
is the pion decay constant in our conventions. 
The light quark electric charge matrix, 
$Q$, 
is given by
$Q = \diag ( 2/3, -1/3)$.

In determining electromagnetic properties at NLO in the 
$SU(2)$ 
chiral expansion, 
there are additional electromagnetic tree-level operators needed from the 
higher-order Lagrangian. 
We give a general discussion of these operators for an arbitrary 
$SU(2)$ 
multiplet before writing down the operators special to the case of the nucleon and delta fields. 
These operators can be grouped by their multipolarity, $\ell$. 
For the spin one-half baryons, 
the allowed multipoles are
$\ell = 0$ and $1$, 
corresponding to the electric charge and magnetic dipole form factors, respectively. 
For the spin three-half baryons, 
the possible multipoles are
$\ell = 0$ -- $3$; 
which, in order,
correspond to the electric charge, 
magnetic dipole, 
electric quadrupole, 
and magnetic octupole
form factors. 
The local operators for each multipolarity
consist of a tower of terms that can be organized
by the number of derivatives. 
Terms possessing more derivatives are of course
higher-order in the power counting. 
Given that the total charge is fixed by gauge invariance, 
the leading electric form factor operators contribute to  the electric 
charge radii. 
These operators contribute at NLO,
and are needed for our calculation. 
For the magnetic form factor, 
the leading magnetic moment operators are LO in the chiral expansion 
and will also be needed for our computation. 
The leading magnetic radii operators, 
however, 
occur at NNLO, and will not be considered here. 
Electric quadrupole operators first appear at NLO in the chiral expansion,
and are hence required in our computation. 
Local contributions to quadrupole radii enter at NNNLO, 
and will not be needed. 
Magnetic octupole moments are not generated at NLO from loops or local contributions. 
Lastly
there are generally two flavor structures permitted for each multipole operator in the limit of strong isospin. 
Linear combinations of these structures can be identified with isovector and isoscalar contributions from the 
electromagnetic current.

Having described in general the types of operators needed, 
let us now give operators relevant for the $S=0$ isospin multiplets. 
The electromagnetic operators for the nucleon have the form:
$O^{Na}_\ell
=
\ol N \cO_\ell \, Q N$, 
and
$O^{Nb}_\ell
=
\ol N \cO_\ell \, N \, \tr  (Q )$,
where only the flavor dependence has been written explicitly.%
\footnote{
The full chiral structure of all local electromagnetic operators
can be obtained by the replacement,
$Q \to \frac{1}{2} (\xi Q \xi^\dagger + \xi^\dagger Q \xi)$,
which generates pion loops at higher orders than we are considering. 
} 
The 
$\cO_\ell$
arise from the multipole expansion of the electromagnetic field,
and are flavor singlets. 
Here the 
$a$ 
and 
$b$ 
merely denote the two possible flavor contractions. 
For 
$\ell = 0$,
we require only the leading contribution to 
$\cO_{\ell =0}$
which gives rise to the electric charge radius,
namely
$\cO_{\ell = 0}
=
e \, v_\mu \partial_\nu F^{\mu \nu} 
$.
The coefficients of the operators 
$O_{\ell =0}^{Na}$ 
and 
$O_{\ell =0}^{Nb}$
are the relevant low-energy constants for the computation of the nucleon charge radii. 
For 
$\ell = 1$, 
we require the leading magnetic moment operators, 
for which the required multipole structure involving the photon field is
$\cO_{\ell = 1}
= 
\frac{i e}{2 M_N}
[S_\mu, S_\nu] F^{\mu \nu}
$.
There are again two low-energy constants corresponding to the coefficients of the 
operators
$O_{\ell =1}^{Na}$ 
and 
$O_{\ell =1}^{Nb}$.

The multipole operators for the delta have a different structure due to the vector indices carried by the Rarita-Schwinger fields. 
For a general multipole operator, 
$\cO_\ell^{\mu \nu}$, 
there are only two flavor contractions possible for the delta, 
namely 
$O^{Ta}_\ell
= 
\left(
\ol T_\mu \, \cO^{\mu \nu}_{\ell} \, Q \, T_\nu
\right)
$,
and
$O^{Tb}_\ell
=
\left(
\ol T_\mu \, \cO^{\mu \nu}_{\ell} \,  T_\nu 
\right)
\, \tr ( Q ) 
$.
Again 
$a$ 
and 
$b$ 
merely denote the two different flavor structures, 
the 
$\cO_\ell^{\mu \nu}$
are flavor singlets, 
and the coefficients of such operators are the required low-energy constants. 
For the electric form factor, we require operators contributing to the charge radius,
for which the relevant photon operator is  
$\cO_{\ell = 0}^{\mu \nu} 
= 
e \,
v_\alpha \partial_\beta F^{\alpha \beta}
g^{\mu \nu} 
$.
Operators contributing to the magnetic moments require 
$\cO_{\ell = 1}^{\mu \nu}
=
\frac{i e}{2 M_N} F^{\mu \nu}
$.
The occurrence of the nucleon mass renders the 
delta magnetic moments in units of nuclear magnetons. 
From our discussion above, 
we lastly require the electric quadrupole operators, 
for which the relevant multipole structure is given by
$\cO_{\ell = 2}^{\mu \nu}
= 
e\, v_\alpha 
\left( 
\partial^{\mu} F^{\a \nu} 
+ 
\partial^{\nu} F^{\a \mu}  
-  
\frac{1}{2}
g^{\mu \nu}
\partial_\beta
F^{\a \beta} 
\right)
$.

\subsection{Strangeness $S=1$ Baryons \label{S=1L}}

At leading order, 
the heavy baryon effective Lagrangian for strangeness $S = 1$ hyperons, 
namely the 
$\L$, 
$\S$, 
and 
$\Sigma^{*}$, 
is given by
\begin{eqnarray} \label{eq:LS=1}
\cL^{(S=1)}_2 &=&
	\ol \L \left( i v \cdot \partial \right) \L
		+ \tr \left[ \ol \S \left( iv \cdot \cD - \D_{\L \S} \right) \S \right]
			- \Big( \ol \S {}^{*\mu} \left[ i v \cdot \cD - \D_{\L \S^*} \right] \S^*_\mu \Big)
,\end{eqnarray}
where the spin one-half
$\S$ 
and spin three-half 
$\S^{*}$
fields can be written in matrix form
\begin{equation}
\S
= 
\begin{pmatrix}
\frac{1}{\sqrt{2}} \S^0  &
\S^+ \\
\S^- &
- \frac{1}{\sqrt{2}} \S^0
\end{pmatrix}, 
\quad
\text{and}
\quad
\S^*
= 
\begin{pmatrix}
\Sigma^{*+} &  \frac{1}{\sqrt{2}} \S^{*0} \\
\frac{1}{\sqrt{2}} \S^{*0} &  \S^{*-}
\end{pmatrix}
.\end{equation}
The former field transforms as an adjoint, 
while the latter transforms as a two-index symmetric tensor. 
Appearing in the free Lagrangian is the parameter 
$\D_{\L\S}=M_\S^{(0)} - M_\L^{(0)}$ 
($\D_{\L\S^*} = M_{\S^*}^{(0)} - M_\L^{(0)}$)
which is the mass splitting between the 
$\S$  
and 
$\L$ 
($\S^*$ and $\L$)
in the chiral limit. 
In writing Eq.~(\ref{eq:LS=1}), 
we have adopted the power counting scheme 
$\e \sim m_{\pi} / \Lambda_\chi \sim k / \Lambda_\chi$, 
where 
$k$ 
is a typical residual momentum, 
and 
$\L_\chi = 2 \sqrt{2} \pi f$ 
is the chiral symmetry breaking scale. 
We will treat
$\D_{\L \S}  = 77 \, \texttt{MeV}$ 
and 
$\D_{\L \S^*}  = 270 \, \texttt{MeV}$ 
as small parameters 
$\sim \e$ 
in our power counting scheme as well.
The covariant derivatives 
$\cD_\mu$ 
appearing above act on $\S$ and $\S^{*}$ 
as 
\begin{eqnarray}
\cD_\mu \Sigma &=&  \partial_\mu \S + [ V_\mu, \Sigma],   
\\
\left( \cD_\mu \Sigma^*_\nu \right)_{ij}
&=&
\partial_\mu (\Sigma^*_\nu)_{ij}
+ 
(V_{\mu})_{i} {}^k  (\Sigma^*_\nu)_{kj}
+
(V_\mu)_{j} {}^k  (\Sigma^*_\nu)_{ik}
- 
\tr ( V_\mu) \, (\Sigma^*_\nu)_{ij}
.\end{eqnarray}

The leading-order interaction terms between the 
$S=1$ 
baryons and pions are contained in the Lagrangian~%
\cite{Tiburzi:2008bk}
\begin{eqnarray}
\cL^{(S=1)}
&=&
g_{\S\S} 
\tr \left(
\ol \S  S^\mu \left[A_\mu , \S \right] 
\right)
+
2 g_{\S^*\S^*}
\left(
\ol \S {}^{*\mu}  S \cdot A   \S^*_\mu   
\right)
+
g_{\S^*\S}
\left(
\ol \S {}^{*\mu}  A_\mu  \S  
+
\ol \S   A^\mu   \S^*_\mu
\right)
\notag \\
&+&
\sqrt{\frac{2}{3}} g_{\L \S}
\Big[
\tr \left( 
\ol \S \, S \cdot A 
\right) \L
+
\ol \L 
\tr \left(
S \cdot A  \S
\right)
\Big]
+ 
g_{\S^* \L}
\Big[
\left( 
\ol \S {}^{*\mu}   A_\mu 
\right) \L
+
\ol \L 
\left(
A^\mu  \S^*_\mu
\right)
\Big]
\label{eq:LinterS=1}
.\end{eqnarray}
The tensor products between spin one-half and spin three-half baryons 
have been denoted with parentheses, 
and are defined by:
$( \ol \S {}^* A \, \S^* )
=
\ol \S {}^{*ij}  A_{j} {}^k \, \S^*_{ki}$,
$
( \ol \S {}^* A \, \S )
= 
\ol \S {}^{*ij}  A_{j} {}^k \, \S_{k} {}^l \, \epsilon_{li}$,
and
$( \ol \S {}^* A )
= 
\ol \S {}^{*ij}  A_{j} {}^k \, \epsilon_{ki}
$.

Electromagnetic interactions have been included in the 
$S=1$ 
Lagrangian using the vector and axial vector fields of pions 
which are given above in 
Eqs.~\eqref{eq:Vector} and \eqref{eq:Axial}, respectively. 
Additionally there are local electromagnetic interactions
required from the higher-order Lagrangian. 
Using the notation set up in the discussion about the 
$S=0$ 
baryons, 
the operators required in the 
$S=1$ 
sector have the form
$O_\ell^{\S a}
= 
\tr
\left(
\ol \Sigma \cO_{\ell} [ Q, \Sigma] 
\right)$,
$O_\ell^{\S b}
=
\tr
\left(
\ol \Sigma \cO_{\ell}  \Sigma
\right) 
\, 
\tr ( Q)$,
and
$O_{\ell}^{\L b}
=
\ol \L \cO_{\ell}  \L 
\, 
\tr (Q)
$,
for the spin one-half baryons, 
$O_\ell^{\S \L a}
=
\ol \L \, \cO_{\ell}
\, \tr ( Q \S )
+ 
\tr (\ol \S Q) 
\cO_{\ell} 
\, \L
$,
for their transitions, 
and finally
$O_\ell^{\S^* a}
= 
\left(
\ol \Sigma {}^*_{\mu} \, \cO^{\mu \nu}_{\ell}  Q \, \Sigma^*_\nu 
\right)$,
and
$O_\ell^{\S^* b}
=
\left(
\ol \Sigma {}^*_\mu \, \cO^{\mu \nu}_{\ell}  \Sigma^*_\mu
\right) 
\, 
\tr ( Q)
$,
for the spin three-half baryons. 
The required 
$\cO_{\ell}$, 
and 
$\cO_\ell^{\mu \nu}$
have been detailed above. 
All operators are accompanied by low-energy constants; 
and, 
in $SU(2)$ \CPT\, 
coefficients in the $S=0$ sector are unrelated to those in the $S=1$ sector.

\subsection{Strangeness $S=2$ Baryons \label{S=2L}}

The leading-order free Lagrangian for strangeness 
$S=2$ 
cascades, 
namely the spin one-half 
$\Xi$ 
and spin three-half 
$\Xi^{*}_{\mu}$, 
is given by
\begin{eqnarray}
\cL^{(S =2)}_2 
&=&
\ol \X \, i v \cdot \cD \, \X 
- 
\ol \X {}^{*\mu} \left( iv \cdot \cD  - \D_{\X \X^*} \right) \X^*_\mu 
\label{eq:LS=2}
\end{eqnarray}
where cascade fields are both packaged as doublets
$\Xi = (\Xi^{0},\Xi^{-})^T$ and 
$\Xi^{\star}_{\mu} = (\Xi^{* 0}_{\mu},\Xi^{* -}_{\mu})^T$.
Above the covariant derivative 
$\cD_\mu$
acts on both 
$\Xi$ 
and 
$\Xi^{*}_{\mu}$ 
in the same manner,
\begin{eqnarray}
(\cD_\mu \Xi )_i 
&=& 
\partial_\mu \Xi_i
+ 
(V_\mu)_{i} {}^{j} \, \Xi_j
- 
2 \, \tr ( V_\mu) \, \Xi_i
\\
(\cD_\mu \Xi^*_\nu )_i 
&=& 
\partial_\mu (\Xi^*_\nu)_i
+ 
(V_\mu)_{i} {}^{j} (\Xi^*_\nu)_j
- 
2 \, \tr ( V_\mu) \, (\Xi^*_\nu)_i
.\end{eqnarray}
Further, 
the parameter 
$\D_{\X \X^*}$ 
is the mass splitting between the 
$\X^*$ 
and 
$\X$ 
in the chiral limit, 
$\D_{\X \X^*} = M_{\Xi^*}^{(0)} - M_{\Xi}^{(0)} = 215\,\texttt{MeV}$, 
and will be treated as a small parameter 
$\sim \e$
in the power counting. 
Additionally
the leading order interaction Lagrangian between the cascade baryons and pions
reads~%
\cite{Tiburzi:2008bk}
\begin{equation}\label{eq:LinterS=2}
\cL^{(S =2)}
=
2 g_{\X\X} \, 
\ol \X  \, 
S \cdot A \,
\X
+ 
2 g_{\X^* \X^*}
\,
\ol \X {}^{*\mu} \, 
S \cdot A \, 
\X^*_\mu
+
g_{\X^* \X}
\left(
\ol \X {}^{*\mu} \,
A_\mu \,
\X
+ 
\ol \X \,
A^\mu \,
\X^*_\mu
\right)
.\end{equation}

Electromagnetism has been included in the 
$S=2$
baryon Lagriangian via the vector and axial 
fields of pions, Eqs.~\eqref{eq:Vector} and \eqref{eq:Axial}. 
Local electromagnetic operators are further required.
In our notation, 
the required operators have the form
$O_\ell^{\X a}
=
\ol \Xi \, \cO_\ell \, Q  \, \X$,
and 
$O_\ell^{\X b}
=
\ol \X \, \cO_\ell \, \X 
\,
\tr ( Q)
$,
for the spin one-half cascades, 
and
$O_\ell^{\X a}
=
\ol \Xi  {}^*_\mu \, \cO_\ell^{\mu \nu} \, Q  \, \X^*_\nu$, 
and
$O_\ell^{\X^* b}
=
\ol \X {}^*_\mu \, \cO^{\mu \nu}_\ell \, \X^*_\nu 
\,
\tr ( Q)
$,
for the spin three-half cascades. 
The required low-energy constants are the coefficients of these operators.

\subsection{Strangeness $S=3$ Baryon \label{S=3L}}

The strangeness $S=3$ baryon is the 
$\Omega$
which is an 
$SU(2)$
singlet. 
The free Lagrangian at leading order is simply
$\cL
=
\ol \Omega {}^\mu \, i  v \cdot \cD \, \Omega_\mu
$,
where the action of the covariant derivative is specified by
$\cD_\mu \Omega_\nu = \partial_\mu \Omega_\nu - 3 \, \tr (V_\mu) \, \Omega_\nu
$,
and only yields a total charge coupling to the photon. 
At this order, 
there are no pion-omega axial interactions~%
\cite{Tiburzi:2008bk}. 
The leading pion interactions arise from 
chiral symmetry breaking operators in the 
$S=3$ 
sector, and generate tadpole graphs which 
scale as $m_\pi^2 \sim \e^2$.
Consequently the electromagnetic properties of the 
$\Omega$ 
are determined by local operators. 
These operators have the form 
$O_\ell^{\O b}
=
\ol \Omega_\mu \cO_\ell^{\mu \nu} \Omega_\nu 
\, \tr ( Q) 
$.

\section{Baryon Electromagnetic Properties}
\label{spin-1/2}

\begin{figure}
\begin{center}
\includegraphics[width=0.5\textwidth]{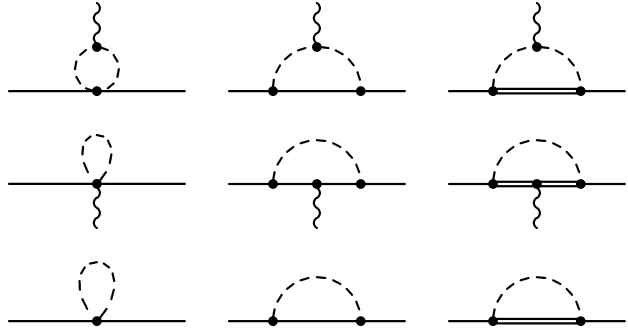}
\end{center}
\caption{One-loop diagrams which contribute at NLO to the 
electromagnetic form factors of spin one-half hyperons. 
Pions are represented by a dashed line, the wiggly line denotes the photon 
and the single (double) lines are the symbols for spin one-half (spin three-half) hyperons. 
Diagrams shown on the bottom row are needed for wavefunction renormalization. 
}
\label{fig0}
\end{figure}

Baryon electromagnetic form factors at or near zero momentum transfer
enable one to extract the electromagnetic moments and radii. 
In the heavy baryon formalism, 
these form factors can be obtained from the matrix
elements of the electromagnetic current
$J_{\mu}$.
For the case of spin one-half baryons, one has the decomposition
\begin{equation}
  \langle\ol B(p') \left|J^\mu\right|B(p)\rangle
  =
 e\,  u^\dagger
  \left\{
    v^\mu G_{E0}(q^2)+\frac{[S^\mu,S^\nu]}{M_N} q_\nu \,  G_{M1}(q^2)
  \right\}
  u,
\end{equation}
where 
$q=p'-p$ 
is the momentum transfer, and 
$u^\dagger$ 
and 
$u$ 
are Pauli spinors. 
The Sachs electric and magnetic form factors are
$G_{E0}(q^2)$ 
and 
$G_{M1}(q^2)$, 
respectively. 
The charge form factor is normalized to the total hadron charge at zero momentum transfer, 
$G_{E0}(0) = Q$. 
The charge radius 
$<r_{E0}^2>$, 
magnetic moment 
$\mu$, 
and magnetic radius
$<r_{M1}^2>$
are defined in terms of the form factors by
\begin{eqnarray} \label{eq:spin1/2}
<r_{E0}^2>
=
6
\frac{d}{dq^2}
G_{E0}(0),
\quad 
\mu=G_{M1}(0), 
\quad
\text{and}
\quad
<r_{M1}^2>
=
6
\frac{d}{dq^2}
G_{M1}(0)
.\end{eqnarray}
Notice that for the charge and magnetic radii, 
we do not employ the physically correct definitions.
We use (six times) the slope of the electric 
and magnetic form factors at zero momentum transfer, 
rather than additionally dividing by the value of the respective
form factors at zero momentum transfer.%
\footnote{
The exception is the definition of electric radii of neutral particles, 
for which one customarily chooses to divide by the proton charge.
} 
We choose to work with the form factor slopes rather than the 
physical radii for two reasons both of which are related to lattice QCD applications. 
Firstly, 
the definitions in Eq.~\eqref{eq:spin1/2} are directly proportional to the current matrix element. 
Differences of current matrix elements satisfy useful properties, 
most notably differences within an isospin multiplet are independent of sea quark charges%
~\cite{Tiburzi:2009yd}. 
Currently lattice QCD computations are largely restricted to vanishing sea quark charges, 
and we have chosen our normalization to expedite comparison with lattice data. 
Secondly, 
in the case of magnetic radii, 
the physical normalization becomes ambiguous in comparing with lattice QCD computations. 
One can choose to divide by either the magnetic moment in nature, or as obtained on the lattice. 
The former does not introduce additional pion mass dependence, 
whereas the latter is a more physical definition. 
We sidestep these issues altogether by using Eq.~\eqref{eq:spin1/2}
and reminding the reader throughout.

The one-loop diagrams necessary to determine the electric and magnetic 
form factors of spin one-half baryons at NLO are shown in Figure~\ref{fig0}.
There are additionally tree-level diagrams with an electromagnetic 
multipole operator insertion. 
These have been calculated but not depicted. 
Results for spin one-half hyperon properties will be presented below in each strangeness sector.

Similar to the spin one-half baryons, the electromagnetic properties of spin three-half baryons
are encoded in their form factors. 
These form factors can be deduced from current matrix elements 
\begin{equation}
\langle \ol T(p') | J^\rho | T(p) \rangle = - e \,  u^\dagger_\mu \, \mathcal{O}^{\mu \rho \nu} \, u_\nu,
\end{equation}
where 
$u_\mu$ 
is a Pauli spinor-vector satisfying
the Rarita-Schwinger type constraints,
$v \cdot u = 0$, 
and 
$S \cdot u = 0$.
The tensor 
$\mathcal{O}^{\mu \rho \nu}$ 
can be parametrized in terms of four independent form factors~%
\cite{Nozawa:1990gt,Arndt:2003we}
\begin{eqnarray}
\mathcal{O}^{\mu \rho \nu} 
&=& 
g^{\mu \nu} 
\left\{ 
v^\rho G_{E0}(q^2) 
+ 
\frac{[S^\rho,S^\tau] }{M_N} q_\tau  
G_{M1}(q^2)  
\right\}
\notag \\
&&-
\frac{1}{2 M_N^2}
\left( q^\mu q^\nu - \frac{1}{4} g^{\mu \nu} q^2 \right) 
\left\{ v^\rho G_{E2}(q^2) + \frac{[S^\rho,S^\tau]}{M_N} q_\tau G_{M3}(q^2)  \right\}
.\end{eqnarray}
The charge and magnetic form factors lead to the charge radii, 
magnetic moments,
and magnetic radii via Eq.~\eqref{eq:spin1/2}.
The electric quadrupole form factor
$G_{E2}(q^2)$
produces the quadrupole moment 
$\cQ$
and 
quadrupole radius
$<r_{E2}^2>$, 
namely
\begin{equation}
\cQ = G_{E2}(0)
, \quad
<r_{E2}^2> 
= 
6 \frac{d}{d q^2} G_{E2}(0)
.\end{equation}
The magnetic octupole moment and radius can be defined similarly. 
Notice again that our definitions do not correspond to the physical electromagnetic radii. 
Instead,
we use (six times) the slope of the form factors at vanishing momentum transfer.

\begin{figure}
\begin{center}
\includegraphics[width=0.5\textwidth]{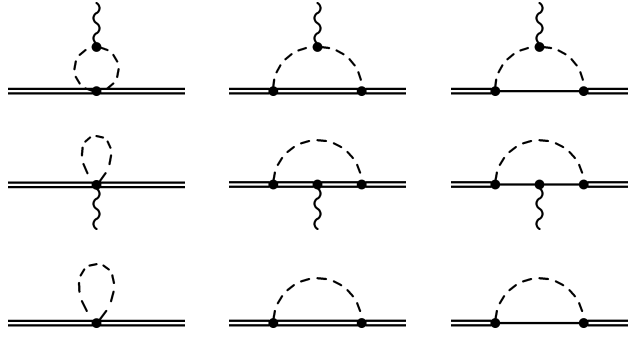}
\end{center}
\caption{One-loop diagrams which contribute at NLO to the electromagnetic 
form factors of spin three-half hyperons. 
At the bottom are wavefunction renormalization diagrams. 
Diagram elements are as in Figure~\ref{fig0}. 
}
\label{fig1}
\end{figure}

To calculate the spin three-half baryon electromagnetic form factors at NLO in HB\CPT, 
we determine the loop diagrams shown in Figure~\ref{fig1}, 
and the tree-level contributions that arise from the insertion 
of LO and NLO electromagnetic operators. 
NLO contributions to the magnetic octupole form factor vanish
leaving only the result for a point-like spin three-half particle, 
which in our units is  
$G_{M3}(0) = Q \, (M_N / M_T)^3$. 
Results for spin three-half hyperon properties will be presented below in each strangeness sector.

\subsection{$S=0$ Baryon Electromagnetic Properties}

Here we collect the results for the nucleon and delta electromagnetic properties at NLO in 
$SU(2)$ HB\CPT. 
The results for the nucleon are not new, 
while a few of the results for the deltas are. 
In this section, 
we furthermore give the definitions of the non-analytic functions arising 
from loop graphs. 
These functions will appear for all baryon electromagnetic properties.

\subsubsection{Nucleons}

The nucleon charge radii are given by~\cite{Bernard:1992qa,Bernard:1998gv}
\begin{eqnarray}
<r_{E0}^2>_{N} 
&=& 
c_0 
+
I_3^{(1/2)} 
c_1(\mu)
-
I_3^{(1/2)}
\frac{2}{(4 \pi f)^2} 
\left[
( 1 + 5g_A^2 )
\log \frac{m_\pi^2}{\mu^2}
-
\frac{20}{9}
g_{\D N}^2 G(\D)
\right]
.\end{eqnarray}
For the nucleon isodoublet, 
the third component of isospin is given by 
$I_3^{(1/2)} = \diag (1/2 , -1/2)$.
At this order, 
the charge radii have non-analytic dependence on the quark masses. 
Contributions with intermediate state deltas
depend upon the function
$G(\d)$,
with the pion mass dependence kept implicit. 
This function has the form
\begin{equation} \label{eq:Gfunc}
G(\d) 
\equiv
G(\d, m_\pi)
=
\log 
\left(
\frac{m_\pi^2}{4 \d^2}
\right)
-
\frac{\d} {\sqrt{\d^2 - m_\pi^2}}
\log
\left(
\frac{\d - \sqrt{\d^2 - m_\pi^2 + i \varepsilon}}{\d + \sqrt{\d^2 - m_\pi^2 + i \varepsilon}}
\right)
.\end{equation}
We have renormalized the function 
$G$ 
so that it satisfies the property,
$G(\d, m_\pi = 0) = 0$. 
The radii depend on two low-energy constants, 
$c_0$ 
and 
$c_1(\mu)$,
which contribute to the isoscalar and isovector radii, respectively.
Only the latter has scale dependence.

The nucleon magnetic moments at NLO take the following form~\cite{Bernard:1992qa,Bernard:1998gv}
\begin{eqnarray}
\mu_{N} 
&=& 
\ol \mu_0
+ 
I_3^{(1/2)} \,
\ol \mu_1
- 
I_3^{(1/2)}
\frac{8 M_N}{(4 \pi f)^2} 
\left[
g_A^2
F(0)
+
\frac{2}{9}
g_{\D N}^2 F(\D)
\right]
,\end{eqnarray}
and depend upon a different function, 
$F(\d)$, 
that is non-analytic in the quark mass.
This function also treats the pion mass dependence implicitly, 
and its explicit form is given by
\begin{eqnarray} \label{eq:Ffunc}
F(\d) 
\equiv
F(\d, m_\pi)
&=&
- 
\d 
\log 
\left(
\frac{m_\pi^2}{4 \d^2}
\right)
+ 
\sqrt{\d^2 - m_\pi^2}
\log
\left(
\frac{\d - \sqrt{\d^2 - m_\pi^2 + i \varepsilon}}{\d + \sqrt{\d^2 - m_\pi^2 + i \varepsilon}}
\right)
,\end{eqnarray}
Notice that we have renormalized 
$F$ 
to satisfy the property,
$F(\d, m_\pi = 0) = 0$.
Consequently the scale dependence of the low-energy constants is exactly cancelled, 
and the renormalized parameters, $\ol \mu_0$ and $\ol \mu_1$, are scale independent.  
A useful value of the function is at $\d = 0$, 
for which we have
$F(0) = \pi m_\pi$.

The magnetic radii of the nucleon doublet take the form~\cite{Bernard:1992qa,Bernard:1998gv}
\begin{equation}
<r_{M1}^2>_N
=
- I_3^{(1/2)}
\frac{8 M_N}{(4 \pi f)^2} 
\left[
g_A^2
\mathcal{F}(0)
+
\frac{2}{9}
g_{\D N}^2 
\mathcal{F}(\D)
\right], 
\end{equation}
with the non-analytic quark mass appearing in the function 
$\mathcal{F}(\d)$,
which is defined as
\begin{equation} \label{eq:cFfunc}
\mathcal{F} (\d) 
\equiv
\mathcal{F}(\d, m_\pi)
= 
\frac{1}{2 \sqrt{\d^2 - m_\pi^2}}
\log
\left(
\frac{\d - \sqrt{\d^2 - m_\pi^2 + i \varepsilon}}{\d + \sqrt{\d^2 - m_\pi^2 + i \varepsilon}}
\right)
.\end{equation}
This function is singular in the chiral limit.
The value of 
${\cal F}(\d)$ 
at 
$\d =0$
is given by
$\mathcal{F}(0) = - \frac{\pi}{ 2 m_\pi} $.
Notice there are no local terms, and consequently no undetermined parameters.

\subsubsection{Deltas}
For the deltas, 
the charge radii at NLO in the 
$SU(2)$ 
chiral expansion take the form
\begin{eqnarray}
< r_{E0}^2>_{\D} 
&=& 
c_{0,\D}
+ 
I_3^{(3/2)}
c_{1,\D}(\mu)
-
\frac{2 I_3^{(3/2)}}{ (4 \pi f)^2} 
\left[
\frac{1}{81} 
( 81 + 25 g_{\D \D}^2 )
\log \frac{m_\pi^2}{\mu^2}
+ \frac{5}{9}
g_{\D N}^2 G(-\D)
\right],
\notag \\
\end{eqnarray}
with the third component of isospin given by 
$I_3^{(3/2)} = \diag ( 3/2, 1/2, -1/2, -3/2)$.
These results are consistent with those obtained by one-loop matching of the  
$SU(3)$ 
results given in%
~\cite{Arndt:2003we,Tiburzi:2009yd}
onto 
$SU(2)$.
Notice that there are only two independent low-energy constants for the 
quartet of deltas, 
$c_{0,\D}$ and $c_{1,\D}(\mu)$,
which correspond to the isoscalar and isovector contributions, 
respectively.
Only the latter is scale dependent.

The function $G(\d)$ depends non-analytically on the quark mass, and has been given above in Eq.~\eqref{eq:Gfunc}. 
For 
$m_\pi > \D$, 
the deltas are stable and 
their radii are real-valued quantities. 
As the pion mass is lowered down to 
$\D$, 
the function 
$G(-\D)$
becomes singular.
At the point 
$m_\pi = \D$, 
the delta can fluctuate into 
$\pi N$ 
states,
and it appears additionally necessary to treat pion radiation
to arrive at finite values for physical quantities. 
With lattice QCD applications in mind, 
we will restrict our attention to the region
$m_\pi > \D$.

For magnetic moments of the deltas, 
a thorough analysis exists in Ref.~\cite{Pascalutsa:2004je} 
using a modified 
$SU(2)$ 
power counting~\cite{Pascalutsa:2002pi}.
With our power counting, 
we have
\begin{eqnarray}
\mu_{\D} 
&=& 
\ol \mu_{0,\D} + I_3^{(3/2)} \ol \mu_{1,\D}
-
I_3^{(3/2)}
\frac{8 M_N}{ (4 \pi f)^2} 
\left[
\frac{1}{27} g_{\D\D}^2 F(0)
+ 
\frac{1}{6} g_{\D N}^2 F(-\D)
\right] 
.\end{eqnarray}
Again there are only two independent low-energy constants among the four
delta magnetic moments. 
The function 
$F(\d)$ 
has been given above in Eq.~\eqref{eq:Ffunc}.
The value of 
$F( -\D)$ 
is not singular when 
$m_\pi = \D$; 
however, 
when 
$m_\pi < \D$,
one must properly handle the imaginary contribution
to the magnetic amplitude in physically measurable cross sections~\cite{Pascalutsa:2004je}. 
The magnetic radii of the deltas do not depend on any local terms
at NLO. 
These radii are given by
\begin{equation}
<r_{M1}^2>_{\D}
=
-
I_3^{(3/2)}
\frac{8 M_N}{ (4 \pi f)^2} 
\left[
\frac{1}{27} g_{\D\D}^2 \mathcal{F}(0)
+ 
\frac{1}{6} g_{\D N}^2 \mathcal{F}(-\D)
\right].
\end{equation} 
The function 
$\mathcal{F}(\d)$ 
entering the magnetic radii has been given in Eq.~\eqref{eq:cFfunc}. 
The value of 
$\cF(-\D)$ 
is singular at 
$m_\pi = \D$, 
but we restrict our attention to 
$m_\pi > \D$.

The electric quadrupole moments of the deltas take the following form at NLO
\begin{eqnarray}
\mathcal{Q}_{\D} 
&=& 
\ol \cQ_{0,\D} 
+ 
I_3^{(3/2)}
\ol \cQ_{1,\D}(\mu)
+
I_3^{(3/2)}
\frac{8 M_N^2}{ (4 \pi f)^2} 
\left[
-
\frac{2}{81} g_{\D\D}^2 
\log \frac{m_\pi^2}{\mu^2}
+ 
\frac{1}{18} g_{\D N}^2 G(-\D)
\right] 
.\end{eqnarray}
The function 
$G(\delta)$ 
is given in Eq.~\eqref{eq:Gfunc}. 
These expressions agree with those derived using one-loop matching of the
$SU(3)$ 
results of%
~\cite{Butler:1994ej}
onto 
$SU(2)$. 
For the quadrupole moments, 
there are both isoscalar and isovector low-energy constants;
the former is scale independent.

Lastly the electric quadrupole radii are given by
\begin{equation}
<r_{E2}^2>_{\D}
=
I_3^{(3/2)}
\frac{8 M_N^2}{ (4 \pi f)^2} 
\left[ 
-
\frac{2}{81}
g_{\D\D}^2 \mathcal{G}(0)
+ 
\frac{1}{18} 
g_{\D N}^2 \mathcal{G}(-\D)
\right].
\end{equation} 
The required function 
$\mathcal{G}$ 
entering the quadrupole radii is given by
\begin{equation}
\mathcal{G}(\d)
= 
\frac{1}{10}
\left[
\frac{2}{\d^2 - m_\pi^2}
+ 
\frac{\d}{[\d^2 - m_\pi^2]^{3/2}}
\log
\left(
\frac{\d - \sqrt{\d^2 - m_\pi^2 + i \varepsilon}}{\d + \sqrt{\d^2 - m_\pi^2 + i \varepsilon}}
\right)
\right]
,\end{equation} 
which satisfies $\mathcal{G}(0) = - (5 m_\pi^2)^{-1}$. 
Notice there are no low-energy constants for the quadrupole radii at this order. 
Both quadrupole moments and quadrupole radii become singular when 
$m_\pi = \D$, 
but we consider the region 
$m_\pi > \D$ 
for which the deltas are stable.

\subsection{$S=1$ Baryon Electromagnetic Properties}

Now we collect expressions for the 
$S=1$ 
baryon electromagnetic properties to NLO in 
$SU(2)$ 
HB\CPT. 
The required functions arising from loop graphs are identical to those appearing in the description of
$S=0$ 
baryon properties.

\subsubsection{$\S$ and $\L$ Baryons}

For the triplet of 
$\S$ 
baryons,
we find their charge radii have the form
\begin{eqnarray}
<r_{E0}^2>_{\S} 
&=& 
c_{0,\S} 
+ 
I_3^{(1)} 
c_{1,\S} (\mu)
-
\frac{I_3^{(1)}}{(4 \pi f)^2} 
\left[
\left(
2
+ 
\frac{5}{2}
g_{\S\S} ^2 
\right) 
\log \frac{m_\pi^2}{\mu^2}
\right.
+ 
\left.
\frac{5}{6} 
g_{\L \S}^2 
G(- \D_{\L \S})
+ 
\frac{5}{3} 
g_{\S^* \S}^2 
G(\D_{\S \S^*})
\right], 
\notag \\
\end{eqnarray}
where the third component of isospin is 
$I_3^{(1)} = \diag ( 1, 0, -1)$. 
There are two low-energy constants, 
$c_{0,\S}$, 
and 
$c_{1,\S}(\mu)$.
These expressions cannot be derived directly from matching 
$SU(3)$ 
results onto 
$SU(2)$. 
In the particular limit
$\D_{\L\S} = 0$, 
however, 
we can check our results by using one-loop matching conditions, 
and they agree with%
~\cite{Kubis:1999xb}.

Notice that the one-loop corrections to the charge radius of the
$\S^0$
vanish. 
The same is true of the 
$\L$, 
for which we have
$<r_{E0}^2>_\L
=
c_\L
+ 
\cO(m_\pi)$.
The leading non-analytic quark mass dependence of the 
$\S^0$ 
and 
$\L$ 
charge radii arises at NNLO.
The one-loop corrections to the transition charge radius between 
$\L$ 
and 
$\S^{0}$ 
baryons, 
however,  
are non-vanishing and lead to the result 
\begin{eqnarray}
<r_{E0}^2>_{\S \L} 
&=& 
c_{\S\L}
-
\frac{1}{ (4 \pi f)^2} 
\left[ 
5 
g_{\L \S} g_{\S \S} G( \D_{\L \S} )
+ 
10 
\sqrt{ \frac{2}{3} }  
g_{\L \S^*} g_{\S \S^*}  G(\D_{\L \S^*})
\right].
\end{eqnarray}
This result can be checked against~\cite{Kubis:1999xb} in the limit of 
$\D_{\L \S} = 0$ 
using one-loop matching of 
$SU(3)$ 
onto 
$SU(2)$.

The magnetic moments of the 
$\S$ 
baryons are given by
\begin{eqnarray}
\mu_{\S} 
&=& 
\ol \mu {}_{0,\S}  + \ol \mu {}_{1,\S}
+
I_3^{(1)}
\frac{4 M_N}{(4 \pi f)^2} 
\left[
- 
\frac{1}{2}
g_{\S\S} ^2 
F(0)
- 
\frac{1}{6} 
g_{\L \S}^2 
F(- \D_{\L \S})
+ 
\frac{1}{6} 
g_{\S^* \S}^2 
F(\D_{\S \S^*})
\right] 
.\end{eqnarray}
The non-analytic quark mass dependence of the 
$I_3 = 0$
baryon magnetic moments vanishes at NLO, 
as can be seen for the $\S^0$. 
Additionally we have
$\mu_\L
=
\ol \mu {}_\L
+
\cO(m_\pi^2)
$,
with the first non-analytic dependence of the form
$\sim m_\pi^2 \log m_\pi^2$
entering at NNLO. 
The transition moment between the 
$\S^0$ 
and 
$\L$
receives NLO corrections, 
and is given by
\begin{eqnarray}
\mu_{\S \L} 
= 
\ol \mu {}_{\S \L}
+
\frac{4 M_N}{ (4 \pi f)^2} 
\left[ 
- 
g_{\L \S} g_{\S \S} F( \D_{\L \S} )
+ 
\sqrt{ \frac{2}{3} }  
g_{\L \S^*} g_{\S \S^*}  F(\D_{\L \S^*})
\right].
\end{eqnarray}
When 
$\D_{\L\S} = 0$, 
expressions for 
$S=1$ 
baryon magnetic moments 
agree with those derived from matching the 
$SU(3)$ 
results of~\cite{Jenkins:1992pi} onto 
$SU(2)$.

At NLO, 
the magnetic radii of the 
$\S$ 
are given by
\begin{equation}
< r_{M1}^2 >_{\S}
=
I_3^{(1)}
\frac{4 M_N}{(4 \pi f)^2} 
\left[
- 
\frac{1}{2}
g_{\S\S} ^2 
\cF(0)
- 
\frac{1}{6} 
g_{\L \S}^2 
\cF(- \D_{\L \S})
+ 
\frac{1}{6} 
g_{\S^* \S}^2 
\cF(\D_{\S \S^*})
\right]
,\end{equation}
while that of the 
$\L$ 
is given by
$< r_{M1}^2>_{\L} = 0 + \cO( m_\pi^0)$.
The magnetic transition radius is given by
\begin{equation}
< r_{M1}^2 >_{\S \L}
=
\frac{4 M_N}{ (4 \pi f)^2} 
\left[ 
- 
g_{\L \S} g_{\S \S} \cF( \D_{\L \S} )
+ 
\sqrt{ \frac{2}{3} }  
g_{\L \S^*} g_{\S \S^*}  \cF(\D_{\L \S^*})
\right]
.\end{equation}
Taking $\D_{\L\S} = 0$, 
we can partially check these expressions by carrying out the one-loop matching of 
$SU(3)$
results in%
~\cite{Kubis:1999xb}
onto 
$SU(2)$.

\subsubsection{$\S^*$ Baryons}

For the triplet of 
$\S^{*}$ 
baryons, 
we consider the stability regime in which 
$m_\pi > \D_{\L\S^*}$, 
and all observables are real-valued.
As the pion mass is lowered, 
the electromagnetic radii and quadrupole moments become singular at 
$m_\pi = \D_{\L\S^*}$ 
and 
$\D_{\S\S^*}$.
To work at these pion masses, 
one must treat the effects from pion radiation. 
The magnetic moments of the 
$\S^*$
do not become singular as the pion mass is lowered; 
however, 
the magnetic amplitude becomes complex-valued.

The electric charge radii of the 
$\S^*$ 
are given to NLO by 
\begin{eqnarray}
< r_{E0}^2 >_{\S^{*}} 
&=& 
c_{0,\S^*} 
+
I_3^{(1)} 
c_{1,\S^*}(\mu)
-
\frac{I_3^{(1)}}{ (4 \pi f)^2} 
\left[
\frac{1}{18} 
( 36 + 25 g^2_{\S^* \S^*})
\log \frac{m_\pi^2}{\mu^2}
\right. \nonumber \\
&+& \left. \frac{5}{3} 
g_{\L \S^*}^2 G(- \D_{\L \S^*} ) 
+ \frac{5}{6}
g_{\S \S^*}^2 G( - \D_{\S \S^*} )
\right].
\end{eqnarray}
The magnetic moments have the form
\begin{eqnarray}
\mu_{\S^{*}} 
&=& 
\ol \mu {}_{0,\S^*} 
+
I_3^{(1)}
\ol \mu {}_{1,\S^*} 
-
I_3^{(1)}
\frac{4 M_N}{ (4 \pi f)^2} 
\left[
\frac{1}{6}  g_{\S^* \S^*}^2 F(0)
+
\frac{1}{4} g_{\S^* \S}^2 F(-\D_{\S \S^*})
+
\frac{1}{2} g_{\L \S^*}^2 F(-\D_{\L \S^*})
\right]
,\notag \\
\end{eqnarray}
while the magnetic radii are
\begin{equation}
< r_{M1}^2 >_{\S^{*}} =
-
I_3^{(1)}
\frac{4 M_N}{ (4 \pi f)^2} 
\left[
\frac{1}{6}  g_{\S^* \S^*}^2 \cF(0)
+
\frac{1}{4} g_{\S^* \S}^2 \cF(-\D_{\S \S^*})
+
\frac{1}{2} g_{\L \S^*}^2 \cF(-\D_{\L \S^*})
\right]
.\end{equation}

For the electric quadrupole form factor,
we have a similar pattern
\begin{eqnarray}
\mathcal{Q}_{\S^{*}} 
&=& 
\ol \cQ {}_{0,\S^*} 
+ 
I_3^{(1)}
\ol \cQ {}_{1,\S^*} (\mu)
+
I_3^{(1)}
\frac{4 M_N^2}{ (4 \pi f)^2} 
\left[
-
\frac{1}{9}  g_{\S^* \S^*}^2 
\log \frac{m_\pi^2}{\mu^2}
\right. \notag \\
&+& \left.
\frac{1}{12} g_{\S^* \S}^2 G(-\D_{\S \S^*})
+
\frac{1}{6} g_{\L \S^*}^2 G(-\D_{\L \S^*})
\right],
\end{eqnarray}
for the quadrupole moments, 
and
\begin{equation}
< r_{E2}^2 >_{\S^{*}} 
=
I_3^{(1)}
\frac{4 M_N^2}{ (4 \pi f)^2} 
\left[
-
\frac{1}{9}  g_{\S^* \S^*}^2 
\cG(0)
+
\frac{1}{12} g_{\S^* \S}^2 \cG(-\D_{\S \S^*})
+
\frac{1}{6} g_{\L \S^*}^2 \cG(-\D_{\L \S^*})
\right],
\end{equation}
for the quadrupole radii. 
In the particular limit 
$\D_{\L\S} =0$, 
we have checked the 
$\S^*$ 
results using one-loop matching from 
$SU(3)$ 
calculations to 
$SU(2)$. 
For the dipole and quadrupole moments, 
the 
$SU(3)$
expressions are contained in%
~\cite{Butler:1994ej}, 
while expressions for the charge radii appear in%
~\cite{Arndt:2003we,Tiburzi:2009yd}.
The magnetic and quadrupole radii, 
however, 
have not been determined in 
$SU(3)$.

\subsection{$S=2$ Baryon Electromagnetic Properties}

The electromagnetic properties of the spin one-half and spin three-half cascades are collected in this section. 
The spin one-half are presented first, followed by the spin three-half.

\subsubsection{$\X$ Baryons}

For the isodoublet of spin one-half cascades, 
we have the following results for their charge radii at NLO in 
$SU(2)$ 
HB\CPT
\begin{eqnarray}
<r_{E0}^2>_{\X} 
&=& 
c_{0,\X}
+ 
I_3^{(1/2)}
c_{1,\X}
-
I_3^{(1/2)} 
\frac{2}{(4 \pi f)^2} 
\left[
(1 + 5 g_{\X \X}^2)
\log \frac{m_\pi^2}{\mu^2}
+ 
\frac{10}{3} g_{\X^* \X}^2  
G(\D_{\X \X^*})
\right]
.\end{eqnarray}
The magnetic moments of the spin one-half cascades fall into a similar pattern
\begin{eqnarray}
\mu_{\X} 
&=& 
\ol \mu_{0,\X}
+
I_3^{(1/2)}
\ol \mu_{1,\X}
+
I_3^{(1/2)}
\frac{8 M_N}{(4 \pi f)^2} 
\left[
- g_{\X \X}^2 F(0)
+ \frac{1}{3} g_{\X^* \X}^2  F(\D_{\X \X^*})
\right]
.\end{eqnarray}
The cascade magnetic radii are given by
\begin{equation}
< r_{M1}^2>_{\X} 
=
I_3^{(1/2)}
\frac{8 M_N}{(4 \pi f)^2} 
\left[
- g_{\X \X}^2 \cF(0)
+ \frac{1}{3} g_{\X^* \X}^2  \cF(\D_{\X \X^*})
\right]
.\end{equation}
These expressions can be derived from 
$SU(3)$ 
results in%
~\cite{Jenkins:1992pi,Kubis:1999xb}
by using the one-loop matching conditions to 
$SU(2)$.

\subsubsection{$\X^*$ Baryons}

For the doublet of 
$\X^{*}$ 
baryons, 
we consider the stability regime in which 
$m_\pi > \D_{\X \X^*}$, 
and all observables are real-valued.
As the pion mass is lowered, 
the electromagnetic radii and quadrupole moments become singular at 
$m_\pi = \D_{\X \X^*}$,
and near this point contributions from pion radiation must be considered.  
The magnetic moments of the 
$\X^*$ baryons
do not become singular as the pion mass is lowered; 
however, 
the magnetic amplitude becomes complex-valued for 
$m_\pi < \D_{\X \X^*}$.

For the electric charge radii of the  
$\Xi^{*}$ 
baryons, 
we obtain
\begin{eqnarray}
<r_{E0}^2>_{\X^{*}} 
&=& 
c_{0,\X^*}
+ 
I_3^{(1/2)} 
c_{1,\X^*}
-
I_3^{(1/2)}
\frac{2}{ (4 \pi f)^2} 
\left[ 
\frac{1}{9}
( 9 + 25 g_{\X^* \X^*}^2)
\log \frac{m_\pi^2}{\mu^2}
+
\frac{5}{3} 
g_{\X \X^*}^2 G( - \D_{\X \X^*} ) 
\right].
\notag \\ 
\end{eqnarray}
The magnetic moments of the 
$\Xi^{*}$
baryons take the form
\begin{eqnarray}
\mu_{\X^{*}} 
&=& 
\ol \mu {}_{0,\X^{*}}
+ 
I_3^{(1/2)}
\ol \mu {}_{1,\X^*}
-
I_3^{(1/2)}
\frac{8 M_N}{ (4 \pi f)^2} 
\left[
\frac{1}{3} 
g_{\X^* \X^*}^2  
F(0)
+
\frac{1}{2}
g_{\X^* \X}^2
F(-\D_{\X \X^*})
\right]
,\end{eqnarray} 
while the magnetic radii are
\begin{equation}
<r_{M1}^2>_{\X^{*}}
=
-
I_3^{(1/2)}
\frac{8 M_N}{ (4 \pi f)^2} 
\left[
\frac{1}{3} 
g_{\X^* \X^*}^2  
\cF(0)
+
\frac{1}{2}
g_{\X^* \X}^2
\cF(-\D_{\X \X^*})
\right]
.\end{equation}

The electric quadrupole moments of the 
$\Xi^{*}$
appear at NLO as
\begin{eqnarray}
\mathcal{Q}_{\X^{*}} 
&=&
\ol \cQ_{0,\X^{*}} 
+
I_3^{(1/2)}
\ol \cQ_{1,\X^*}(\mu)
+
I_3^{(1/2)}
\frac{8 M_{N}^2}{ (4 \pi f)^2} 
\left[
-
\frac{2}{9} 
g_{\X^* \X^*}^2  
\log \frac{m_\pi^2}{\mu^2}
+
\frac{1}{6}
g_{\X^* \X}^2
G(-\D_{\X \X^*})
\right] 
,\end{eqnarray}
while the quadrupole radii are 
\begin{equation}
< r_{E2}^2>_{\X^{*}} 
= 
I_3^{(1/2)}
\frac{4 M_{N}^2}{ (4 \pi f)^2} 
\left[
-
\frac{2}{9} 
g_{\X^* \X^*}^2  
\cG(0)
+
\frac{1}{6}
g_{\X^* \X}^2
\cG(-\D_{\X \X^*})
\right]
.\end{equation}
One can verify these expressions by using one-loop matching conditions to 
$SU(2)$
on the 
$SU(3)$ 
results given in 
~\cite{Butler:1994ej,Arndt:2003we,Tiburzi:2009yd}.

\subsection{$S=3$ Baryon Electromagnetic Properties}

Finally for the isosinglet $\Omega$,
the non-analytic quark mass dependence vanishes at NLO.  
The leading quark mass dependence of its
electromagnetic observables is entirely analytic.
Specifically, each electromagnetic observable 
$\cO_\O$
can be written in the form
$\cO_\O = \alpha_\cO + \beta_\cO m_\pi^2$.

\section{Discussion}
\label{discussion}

Above we have derived expressions for the various electromagnetic properties of hyperons in 
$SU(2)$ 
\CPT. 
To explore the behavior of these properties in two-flavor
chiral expansion, 
we consider two aspects. 
First we investigate the efficacy of the two-flavor expansion by considering
the contributions from virtual kaons. 
Next we estimate the size of 
$SU(2)$ 
chiral corrections by using phenomenological input to determine
the low-energy constants of the two-flavor theory. 
Here we also explore the pion mass dependence of the electromagnetic
properties, and make contact with available lattice QCD data.

\subsection{Kaon Contributions}
\label{KTh}

The spin three-half hyperon resonances are not considerably far from inelastic thresholds. 
For example, 
the 
$\S^*$ 
resonance lies a mere 
$0.05 \, \texttt{GeV}$
below threshold for 
$K N$ 
decay. 
It is natural to wonder how well the non-analyticies associated with 
kaon production are described in a two-flavor expansion. 
Consider a generic 
$\D S = -1$
strangeness-changing baryon transition, 
$B' \to K B$. 
The 
$SU(3)$-breaking 
mass splitting between baryons we denote
$\d_{BB'}$, 
and is given by
\begin{equation}
\d_{BB'} = M_{B'} - M_B
.\end{equation}
When 
$\d_{BB'} > m_K$, 
the decay is kinematically allowed. 
While none of the hyperons lie above the kaon production threshold, 
some are not considerably far below, 
such as the 
$\S^*$. 
In 
$SU(2)$
\CPT, 
the relevant expansion parameter describing virtual kaon contributions has been determined~%
\cite{Tiburzi:2009ab,Jiang:2009fa}
\begin{equation}
\varepsilon_{BB'}
= 
\frac{\frac{1}{2} m_\pi^2}{\frac{1}{2} m_{\eta_s}^2 - \delta_{BB'}^2}
,\end{equation}
where
$m_{\eta_s}$
is the mass of the pseudoscalar $s \ol s$ meson. 
While not a physically propagating particle, 
the 
$\eta_s$
mass can be determined using \CPT, 
or calculated using lattice QCD. 
The latter yields the value
$m_{\eta_s} = 0.686 \, \texttt{GeV}$
~\cite{Davies:2009ts,Aoki:2009ix}.
Breakdown of the 
$SU(2)$
description is possible due to the pole in the expansion parameter, 
$\varepsilon_{BB'}$. 
The 
$SU(2)$
expansion of the virtual kaon contributions, 
however,
is expected to behave reasonably because of the size of 
$\varepsilon_{BB'}$
at the physical pion mass. 
For the worst case scenario, 
we have the largest expansion parameter
$\varepsilon_{N\S^*} = 0.24$,
although higher-order corrections shift this value upwards. 
Not all low-energy observables are safe, however. 
Processes with external momentum, 
for example 
$\pi \S^*$ 
scattering, 
have not been considered, 
and certainly must fail
in 
$SU(2)$ \CPT\ 
above 
$0.05 \, \texttt{GeV}$. 
For the low-energy properties determined in this work, 
we explore kaon contributions on an observable-by-observable basis.

In 
$SU(3)$
\CPT\, 
the kaon loop diagrams with an
$SU(3)$-breaking
baryon mass splitting
generically involve a logarithm 
depending on both 
$m_K$ 
and 
$\d_{BB'}$.
For diagrams of the sunset type, 
the logarithm has the form
\begin{equation}
\cL
(m_K^2, - \d_{BB'} ) 
=
\log
\left(
\frac{- \d_{BB'} - \sqrt{\d_{BB'}^2 - m_K^2 + i \epsilon}}{- \d_{BB'} + \sqrt{\d_{BB'}^2 - m_K^2 + i \epsilon}}
\right)
,\end{equation}
and contains the non-analyticites associated with kaon production 
(which occurs when 
$\d_{BB'} > m_K$). 
Virtual kaon contributions to baryon magnetic moments, 
for example, 
are described by the function
\begin{equation}
F_{Th}(m_K^2, - \d_{BB'} )
= 
\left( \d_{BB'}^2 - m_K^2 \right)^{1/2} 
\cL ( m_K^2,  - \d_{BB'} ) 
,\end{equation}
where we have retained only the non-analyticites that can be associated with kaon production.
The omitted chiral logarithm, 
$\log m_K^2$,
has a well-behaved expansion about the
$SU(2)$ 
chiral limit, 
with an expansion parameter, 
$\varepsilon_{SU(2)} = m_\pi^2 / m_{\eta_s}^2 = 0.04$. 
The
$SU(2)$
expansion of 
$F_{Th}(m_K^2, \d_{BB'} )$
is well-behaved for 
$m_\pi \lesssim 0.3 \, \texttt{GeV}$. 
This was demonstrated in~\cite{Jiang:2009fa}, 
i.e.~the same function enters the description of kaon contributions
to hyperon axial charges. 
Thus we conclude that an 
$SU(2)$
expansion of hyperon magnetic moments 
can describe the non-analytic kaon loop contributions
for both spin one-half and spin three-half hyperons.

%
\begin{figure}[t]
\begin{center}
\epsfig{file=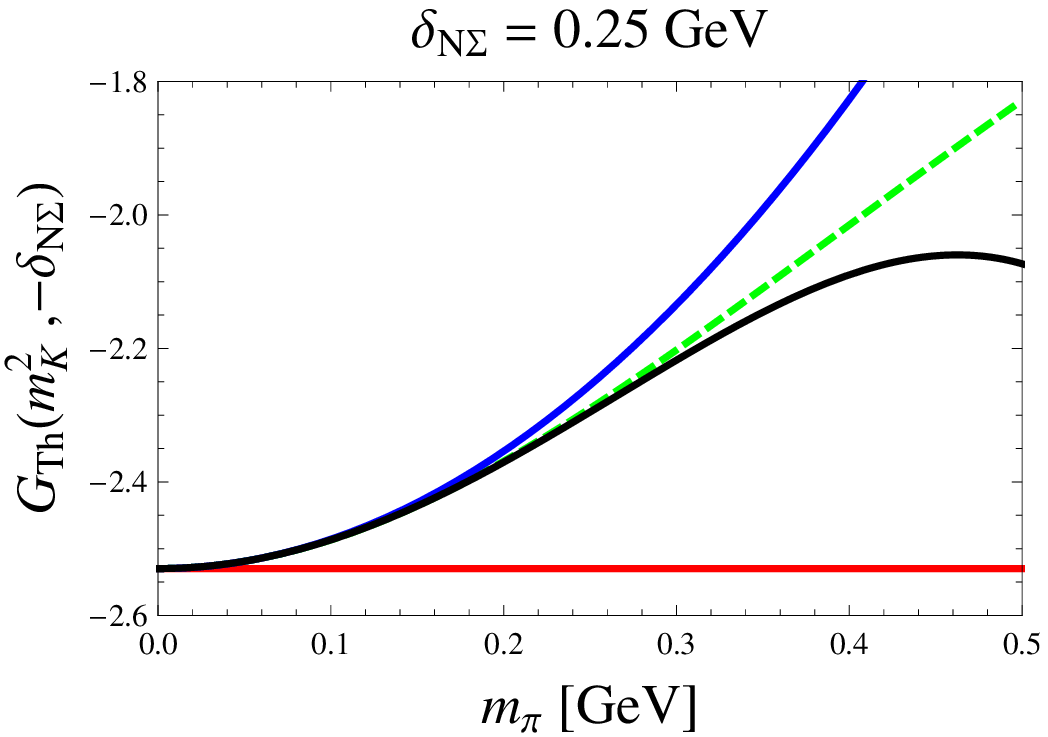,width=1.95in}
$\quad$
\epsfig{file=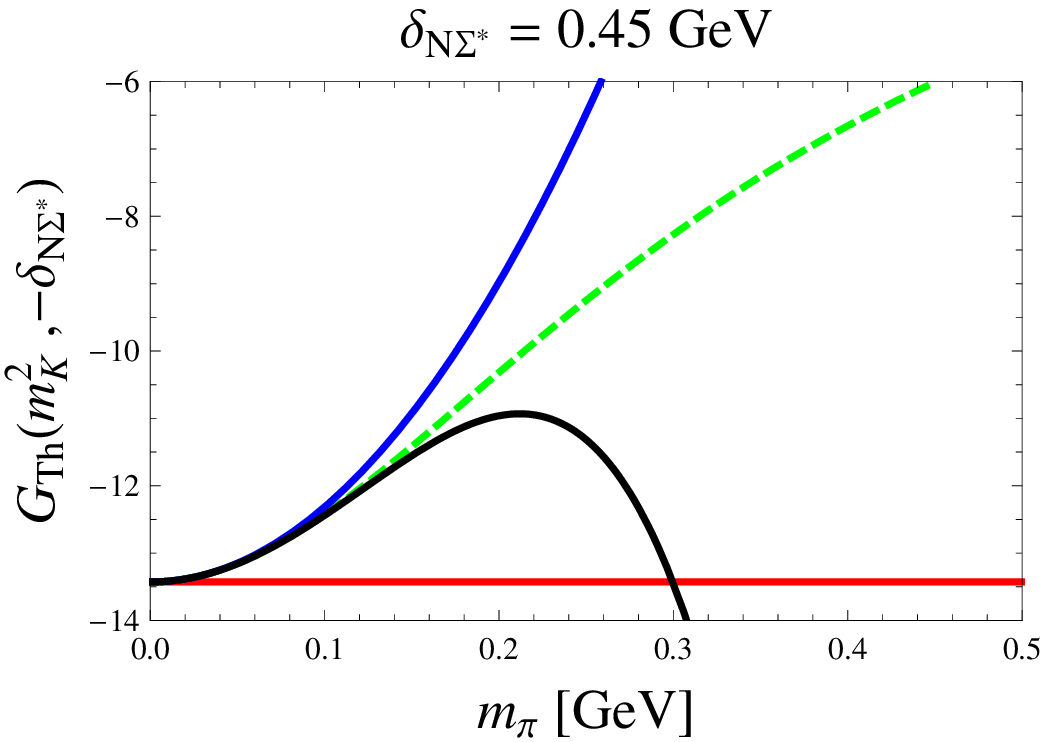,width=1.95in}
$\quad$
\epsfig{file=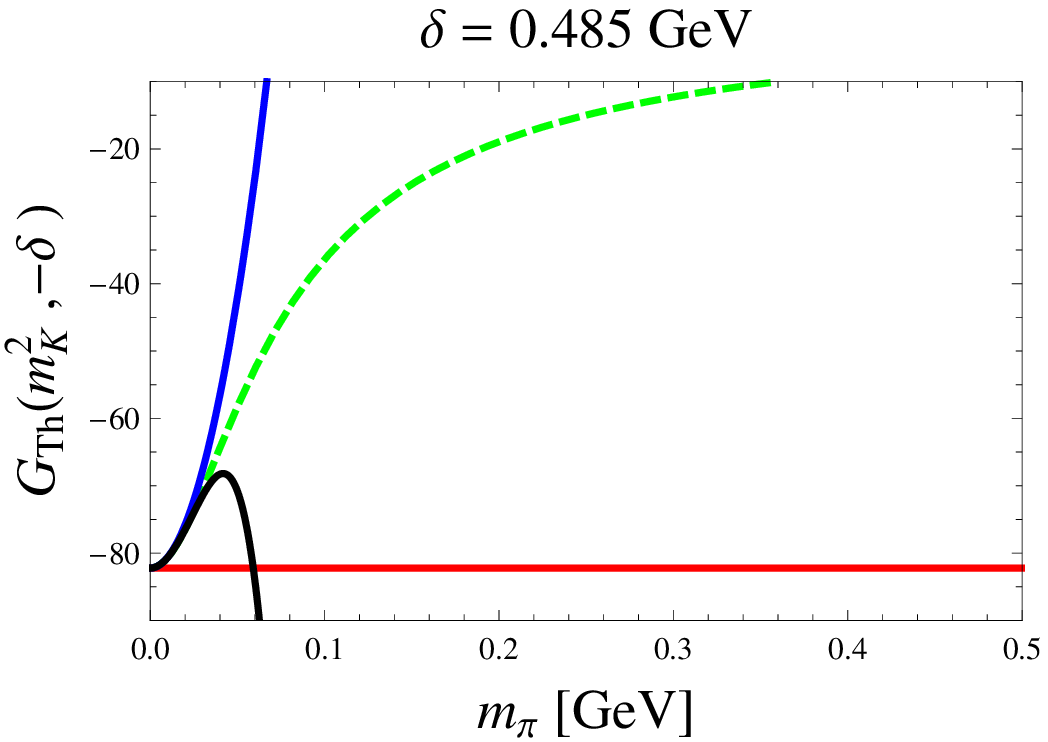,width=1.95in}
\caption{\label{f:Gexpand} 
Contribution from the
$K$-$N$ 
loop diagram for the charge and magnetic radii of 
$\Sigma$ (left) 
and 
$\Sigma^*$ (middle) baryons.
The kaon contribution is also the same for the quadrupole moment of the
$\Sigma^*$. 
Plotted versus the pion mass and shown in dashed green 
is the non-analytic contribution 
$G_{Th}(m^2_K, - \delta_{BB'})$.
Also shown is the loop contribution for a heavier external-state baryon (right)
with splitting 
$\d = 0.485 \, \texttt{GeV}$.  
Compared with these curves are the first three approximations that are analytic in 
$m_\pi^2$, 
see Eq.~\eqref{eq:Gexp}. 
The red curve is the zeroth-order approximation, 
the blue curve includes the first-order correction proportional to 
$m_\pi^2$,
and finally the black curve includes all terms to 
$m_\pi^4$. 
Notice from left to right each plot progressively shows a span of ten greater in range. 
}
\end{center}
\end{figure}
%

The kaon contributions to the remaining hyperon electromagnetic properties must be investigated. 
The electric charge radii, magnetic radii, and electric quadrupole moments
all receive long-distance kaon contributions proportional to the function
\begin{equation} \label{eq:GTh}
G_{Th} (m_K^2, - \d_{BB'} )
=
\frac{\d_{BB'}}
{\left( \d_{BB'}^2 - m_K^2 \right)^{1/2}} 
\,
\cL (m_K^2, - \d_{BB'} )
.\end{equation}
In the 
$SU(2)$
chiral expansion, 
this function is approximated by terms analytic in the pion mass squared, 
namely
\begin{equation} \label{eq:Gexp}
G_{Th} ( m_K^2, - \d_{BB'} ) 
=
G_{Th}^{(0)}
+ 
m_\pi^2 \,
G_{Th}^{(2)}
+ 
m_\pi^4 \,
G_{Th}^{(4)}
+ 
\ldots
\, ,\end{equation}
where only the pion mass dependence has been explicitly shown. 
The first few terms in the expansion are given by
\begin{eqnarray}
G_{Th}^{(0)}
&=&
G_{Th} \Big( \frac{1}{2} m_{\eta_s}^2, - \d_{BB'} \Big),
\notag \\
G_{Th}^{(2)}
&=&
- 
\frac{1}{\d_{BB'}^2 - \frac{1}{2} m_{\eta_s}^2} 
\left(
\frac{\d_{BB'}^2}{m_{\eta_s}^2}
-
\frac{1}{4} G_{Th}^{(0)}
\right),
\notag \\
G_{Th}^{(4)}
&=&
\frac{1}{8}
\frac{1}{ [ \d_{BB'}^2 - \frac{1}{2} m_{\eta_s}^2]^2} 
\left[
\frac{\d_{BB'}^2}{m_{\eta_s}^2}
\left( 
4 \d_{BB'}^2 - 5 m_{\eta_s}^2
\right)
+ 
\frac{3}{4} 
G_{Th}^{(0)}
\right]
.\end{eqnarray}
These terms have non-analytic dependence on the strange quark mass
and constitute the matching conditions between the two- and three-flavor 
theories.

To explore the 
$SU(2)$ 
expansion of kaon contributions to hyperon charge and magnetic radii,
we show the kaon contribution 
$G_{Th}(m_K^2, -\d_{BB'})$ 
of Eq.~\eqref{eq:GTh}
in Fig.~\ref{f:Gexpand}.
Here we specialize to the case 
$K N$ 
fluctuations of the 
$\S$ 
and 
$\S^*$
baryons. 
The depicted 
$K N$ 
contributions are also relevant for the quadrupole moments of 
$\Sigma^*$ 
hyperons. 
The full result is compared with successive approximations derived by expanding about the 
$SU(2)$ 
chiral limit. 
The results show that the virtual kaon contributions can be described 
in the two-flavor effective theory. 
Results are better for the 
$\S$
baryon, 
as the perturbative expansion appears to be under control up to 
$m_\pi \sim 0.3 \, \texttt{GeV}$. 
For the 
$\S^*$, 
however, 
the perturbative expansion does not hold very far beyond the physical pion mass. 
The figure also depicts a fictitious case where the mass splitting takes the value
$\d = 0.485 \, \texttt{GeV}$.
For this splitting, 
the expansion parameter is not small, 
$\varepsilon_{BB'} = 6.9$, 
and 
the range of pion masses for which an 
$SU(2)$
treatment remains effective is exceedingly small.

%
\begin{figure}[t]
\begin{center}
\epsfig{file=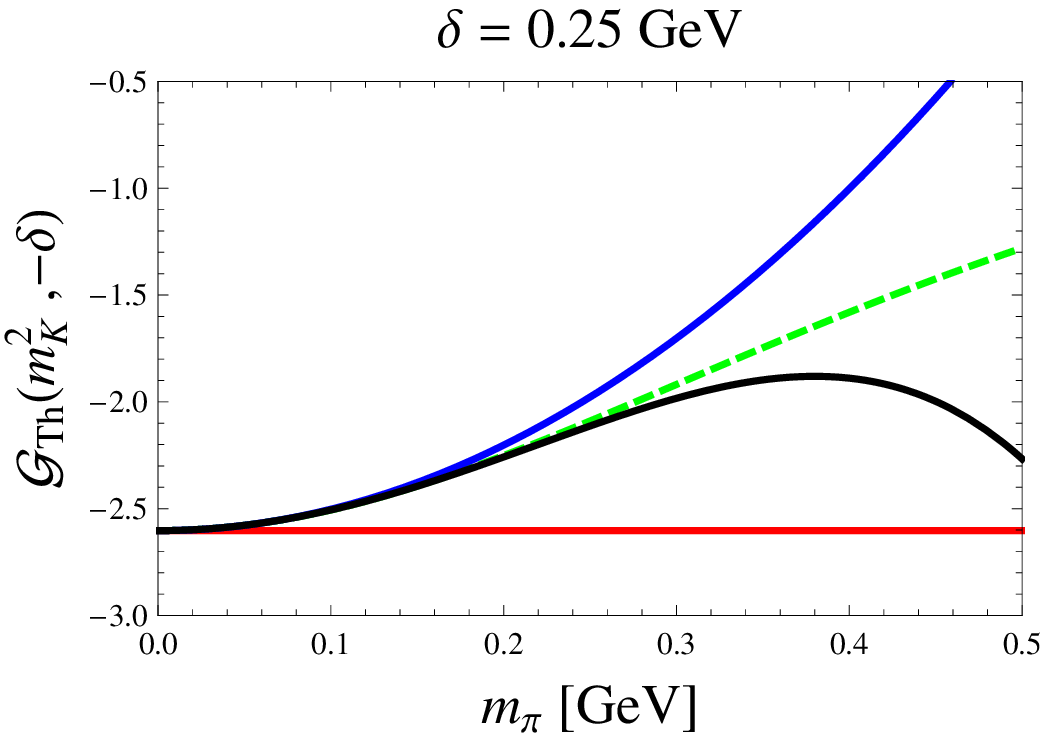,width=1.95in}
$\quad$
\epsfig{file=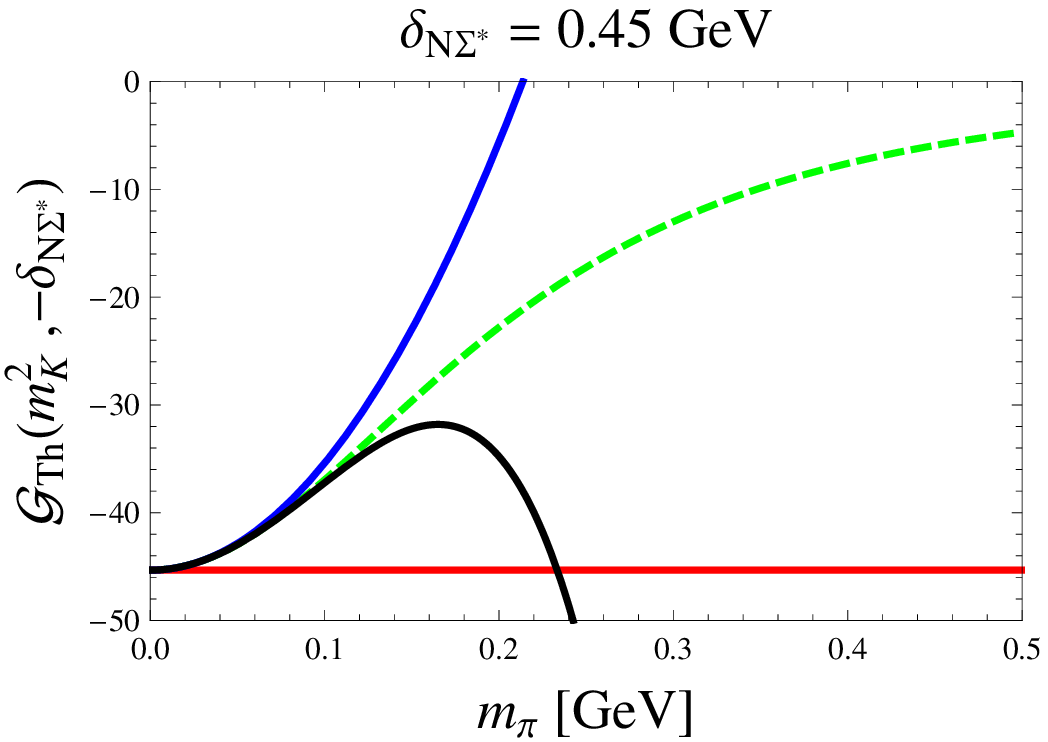,width=1.95in}
$\quad$
\epsfig{file=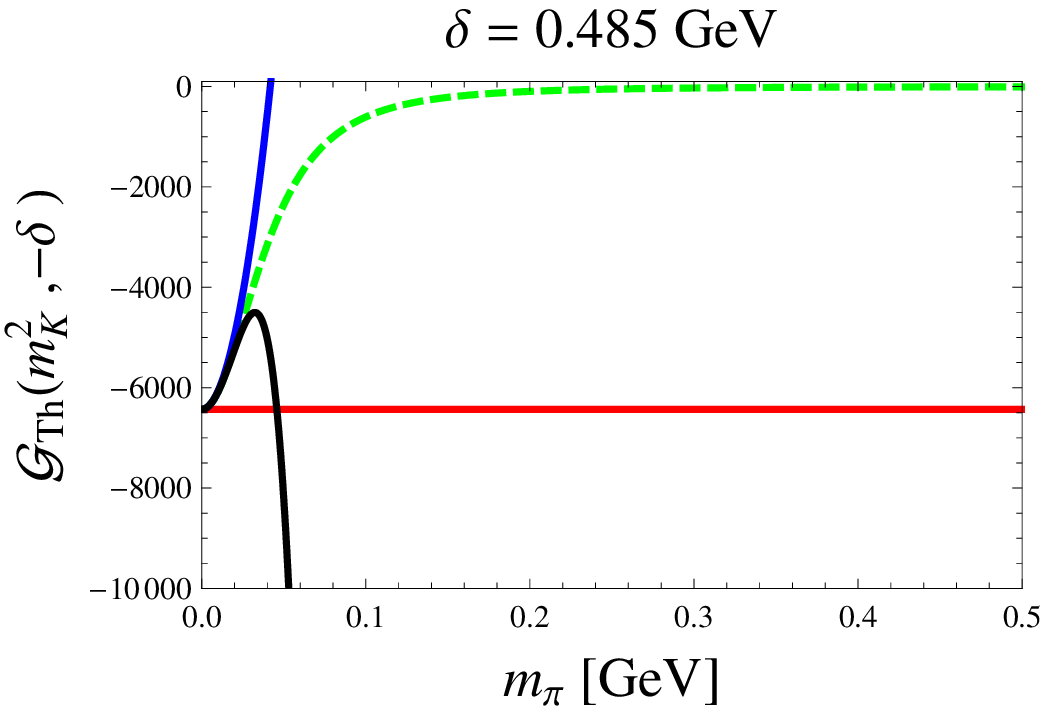,width=1.95in}
\caption{\label{f:GGexpand} 
Contribution from the
$K$-$N$ 
loop diagram for the quadrupole radius. 
Plotted versus the pion mass and shown in dashed green 
is the non-analytic contribution 
$G_{Th}(m^2_K, - \delta_{BB'})$.
We show three plots corresponding to three different mass splittings. 
Results for the $\Sigma^*$ quadrupole radii are shown in the middle.
Also shown is the kaon contribution for a lighter and heavier external-state baryon (left and right)
with splittings 
$\d = 0.25 \, \texttt{GeV}$
and
$\d = 0.485 \, \texttt{GeV}$, 
respectively.  
Compared with these curves are the first three approximations that are analytic in 
$m_\pi^2$, 
see Eq.~\eqref{eq:GGexp}. 
The red curve is the zeroth-order approximation, 
the blue curve includes the first-order correction proportional to 
$m_\pi^2$,
and finally the black curve includes all terms to 
$m_\pi^4$. 
Notice from left to right the plots show a span of twenty and then two-hundred greater in range. 
}
\end{center}
\end{figure}
%

The long-distance kaon contributions to the electric quadrupole radii
are proportional to a different function
\begin{equation} \label{eq:GGTh}
\cG_{Th} (m_K^2, - \d_{BB'} )
=
\frac{1}{10}
\left[
\frac{2}
{ \d_{BB'}^2 - m_K^2 } 
-
\frac{\d_{BB'}}{\left( \d_{BB'}^2 - m_K^2 \right)^{3/2}}
\,
\cL (m_K^2, - \d_{BB'} )
\right]
.\end{equation}
In the 
$SU(2)$
chiral expansion, 
this function is approximated by a series of terms analytic in the pion mass squared, 
\begin{equation} \label{eq:GGexp}
\cG_{Th} ( m_K^2, - \d_{BB'} ) 
=
\cG_{Th}^{(0)}
+ 
m_\pi^2 \,
\cG_{Th}^{(2)}
+ 
m_\pi^4 \,
\cG_{Th}^{(4)}
+ 
\ldots
\, ,\end{equation}
where only the pion mass dependence has been explicitly shown. 
The first few terms in the expansion are given by
\begin{eqnarray}
\cG_{Th}^{(0)}
&=&
\cG_{Th} \Big( \frac{1}{2} m_{\eta_s}^2, - \d_{BB'} \Big),
\notag \\
\cG_{Th}^{(2)}
&=&
\frac{1}{10}
\frac{1}{[\d_{BB'}^2 - \frac{1}{2} m_{\eta_s}^2]^2} 
\left(
1
+
\frac{\d_{BB'}^2}{m_{\eta_s}^2}
- 
\frac{3}{4} G_{Th}^{(0)}
\right),
\notag \\
\cG_{Th}^{(4)}
&=&
\frac{1}{20}
\frac{1}{ [ \d_{BB'}^2 - \frac{1}{2} m_{\eta_s}^2]^3} 
\left[
1
+ 
\frac{9}{4} \frac{\d_{BB'}^2}{m_{\eta_s}^2}
- 
\frac{\d_{BB'}^4}{m_{\eta_s}^4}
-
\frac{15}{16} 
G_{Th}^{(0)}
\right]
.\end{eqnarray}
Notice we have written these latter two expressions in terms of 
$G_{Th}^{(0)}$, 
as opposed to 
$\cG_{Th}^{(0)}$.

To explore the 
$SU(2)$ 
chiral expansion of hyperon electric quadrupole radii, 
we plot the kaon contribution 
$\cG_{Th}( m_K^2, - \d_{BB'} )$
as a function of the pion mass in Fig.~\ref{f:GGexpand}.
Shown along with 
$\cG_{Th}$
are successive approximations to this function that are
derived by expanding
$\cG_{Th}$
 in powers of the pion mass squared. 
We consider three test values for the splitting, 
$\d = 0.25 \, \texttt{GeV}$, 
$\d_{N \S^*} = 0. 45 \, \texttt{GeV}$, 
and 
$\d = 0.485 \, \texttt{GeV}$. 
From Fig.~\ref{f:Gexpand}, 
we see that:
for the smallest splitting the expansion works beyond twice the value of the physical pion mass, 
for the $N \S^*$ splitting the expansion works up to a little beyond the physical pion mass, 
and for the largest splitting the expansion works only for vanishingly small pion masses. 
This pattern is the same as that observed in Fig.~\ref{f:Gexpand}.

From this detailed investigation of kaon loop contributions, 
we expect the magnetic moments of spin one-half and spin three-half 
hyperons to be well described in an expansion about the 
$SU(2)$ 
chiral limit. 
The same is true for the electromagnetic radii of spin one-half hyperons, 
for which we have seen kaon contributions remain perturbative up to 
$m_\pi \sim 0.3 \, \texttt{GeV}$. 
For the radii and quadrupole moments of spin three-half hyperons, 
however, 
the two-flavor chiral expansion is not effective very far beyond the physical pion mass. 
The increased sensitivity in these observables
is due to the threshold singularities in the non-analytic functions, 
Eqs.~\eqref{eq:GTh} and \eqref{eq:GGTh}. 
By contrast, 
kaon contributions to 
the masses, 
axial charges, 
and magnetic moments vanish at the kaon threshold
due to phase-space factors. 
For the case of radii and quadrupole moments, 
the kaon contributions become singular near threshold.
This non-analyticity is not well described by an 
$SU(2)$ 
expansion, 
although the behavior appears to be under control at the physical pion mass. 
Based on this observation, 
we expect curvature terms
(arising from the second derivatives of the form factors)
to be poorly behaved in 
$SU(2)$.

\subsection{$SU(2)$ Chiral Corrections}
\label{SU2ChC}

%
  \begin{table}[t]
  \begin{center}
     \caption{Parameter values for the $SU(2)$ chiral Lagrangian.
      Mass splittings are taken from experiment, 
      while the axial couplings are determined from multiple sources, 
      as described in the text. 
      Listed values for the magnetic moment and charge radius couplings are estimates made using experimental input and the expressions derived from            	$SU(2)$ \CPT\ in this work.  
      In the interests of space, we employ obvious abbreviations for the electromagnetic low-energy constants.  
%
     }
   \smallskip
   \begin{tabular}{c|c|c|c|c}
   Strangeness &
   Mass Splittings $[\texttt{GeV}]$ &
   Axial Charges &
   Magnetic Moments &
   Electric Radii $[\texttt{fm}^2]$\\
   \hline
   \hline
   $S=0$ & 
   $\D = 0.29 $ &
   $g_A =1.27$ &
   $\ol \mu_p = 3.88$ &
   $c_p(\L_\chi) = 0.096$ \\
   $$ & 
   $$ &
   $g_{\D N} = 1.48$ &
   $\ol \mu_n = -3.00$ &
   $c_{n}(\L_\chi) = 0.55$ \\
   $$ & 
   $$ &
   $g_{\D \D} = -2.2$ &
   $\ol \mu_{\D^{++}} = 6.3$ &
   $$ \\
   $$ & 
   $$ &
   $$ &
   $\ol \mu_{\D^{+}} = 2.7$ &
   $$ \\
\hline
   $S=1$ & 
   $\D_{\L \S} = 0.077$ &
   $g_{\L \S} = 1.47$ &
   $\ol \mu_{\S^+} = 2.87$ &
   $c_{\S^-}(\L_\chi) = 0.94$ \\
   $$ & 
   $\D_{\S \S^*} = 0.19$ &
   $g_{\S \S} = 0.78$ &
   $\ol \mu_{\S^-} =  -1.57$ \\
   $$ & 
   $\D_{\L \S^*} = 0.27 $ &
   $g_{\S^* \L} = -0.91$ &
   $\ol \mu {}_{\L} = - 0.613$ &
   $$ \\
   $$ &
   $$ &
   $g_{\S^* \S} = 0.76$ &
   $\ol \mu {}_{\S \L} = 1.90$ &
   $$ \\
   $$ & 
   $$ & 
   $g_{\S^* \S^*} = - 1.47$ &
   $$ &
   $$ \\ 
\hline
  $S=2$ & 
   $\D_{\X \X^*} = 0.215$ &
   $g_{\X \X} = 0.24$ &
   $\ol \mu_{\X^0} = -1.25$ &
   $$ \\
   $$ & 
   $$ &
   $g_{\X^* \X^*} = -0.73$ &
   $\ol \mu_{\X^-} = -0.65$ &
   $$ \\
   $$ & 
   $$ & 
   $g_{\X^* \X} = 0.69$ &
   $$ & 
   $$ \\
\hline
\hline
       \end{tabular}
  \label{t:parameters}
  \end{center}
 \end{table}
%

To investigate the chiral corrections to hyperon electromagnetic properties, 
we use phenomenology to fix the values of the low-energy constants. 
The values of masses, 
magnetic moments, 
and charge radii are taken solely from experiment~\cite{Amsler:2008zzb}. 
For the axial charges, 
we use known experimental values, when available,
and lattice extrapolated values for 
$g_{\S \S}$, 
and 
$g_{\X \X}$~\cite{Lin:2007ap}. 
For the axial charges of spin three-half hyperons,
little information is known, 
and so we adopt the 
$SU(3)$ 
chiral perturbation theory estimate~\cite{Butler:1992pn},
along with tree-level matching conditions between the 
$SU(2)$ 
and 
$SU(3)$ 
theories.\footnote{%
One-loop matching may modify these resonance axial charges considerably, 
as is suggested by considering the loop corrections to the tree-level value of
the axial charge of the delta resonance, $g_{\D\D}$, in $SU(2)$~\cite{Jiang:2008we}.
The axial charges of decuplet baryons, however, have not been calculated beyond tree level in 
$SU(3)$. 
}

From the values of the low-energy constants, 
we can address to what extent loop contributions 
are perturbative. 
The 
$SU(2)$ 
HB\CPT\ 
results show an improvement over 
$SU(3)$
HB\CPT\
for some of the electromagnetic properties. 
For example, 
the size of one-loop corrections to the octet baryon magnetic moments
has been shown in Table~\ref{t:loops}
for both 
$SU(2)$ 
and 
$SU(3)$. 
For the nucleon magnetic moments, 
there appears to be no reason to choose 
$SU(2)$ 
over 
$SU(3)$. 
For the strangeness 
$S=1$ 
hyperons, 
however, 
there is improvement in most cases
and for the 
$S=2$ 
baryons,
the improvement is phenomenal. 
This pattern of improvement follows that seen 
for baryon masses~\cite{Tiburzi:2008bk}, and axial charges~\cite{Jiang:2009sf}. 
There are two transparent physical reasons for improvement with increasing strangeness. 
Firstly the non-relativsitic approximation increases in validity with increasing strangeness. 
Secondly the axial coupling constants generally decrease in size with increasing strangeness. 
Comparing the nucleon and cascade magnetic moments, 
we see that the ratio of pion-cascade loops to pion-nucleon loops scales as
$g_{\X \X}^2 / g_{A}^2 = 0.04$.  
There is a further reduction in the chiral corrections to the cascade magnetic moment
arising from isospin algebra: 
pion-delta quartet loops and pion-cascade resonance loops differ by a sign. 
In the case of the nucleon magnetic moment, both one-loop graphs come with the same sign, 
while in the case of cascade magnetic moments, the two one-loop graphs come with opposite signs. 
This sign difference leads to a cancellation of terms that are already small in magnitude compared to 
the nucleon case.

While the $SU(2)$ theory suffers from a mild proliferation of low-energy constants, 
there are a few quantities for which we can make predictions. 
We are able to determine the magnetic moments%
\footnote{
The imaginary parts for the magnetic moments of deltas and hyperon resonances can also be determined.
}
\begin{eqnarray}
\mu_{\S^0}
&=&
0.65 ,
\qquad
\Re \mathfrak{e} 
\left( \mu_{\D^0} \right)
=
-0.74,
\qquad
\Re \mathfrak{e} 
\left( \mu_{\D^-} \right)
=
-4.2
.\end{eqnarray}
Our value for 
$\mu_{\S^0}$
agrees well with that determined from 
$SU(3)$ 
covariant baryon \CPT\ without decuplet fields~\cite{Geng:2008mf}. 
The value we find for 
$\Re \mathfrak{e} \left( \mu_{\D^-} \right)$
is 
$\sim 20 \%$
smaller than that derived from 
$SU(3)$ covariant baryon \CPT~\cite{Geng:2009ys}.
Corrections arising from NNLO terms [see Eq.~\eqref{eq:AB} below], 
however, 
could easily bring our value into agreement. 
While our value for 
$\Re \mathfrak{e} \left( \mu_{\D^0} \right)$
differs from that of~\cite{Geng:2009ys},
this magnetic moment is small and
NNLO corrections to our result are expected to be comparatively large. 
Our one-loop expressions for magnetic moments exhibit the isospin relations
\begin{eqnarray}
2 \mu_{\S^0}  
&=&
\mu_{\S^+} 
+
\mu_{\S^-},
\qquad
2 \mu_{\S^{*,0}}  
=
\mu_{\S^{*,+}} 
+
\mu_{\S^{*,-}},
\qquad
\mu_{\D^{++}} - \mu_{\D^-} 
=
3 ( \mu_{\D^+} - \mu_{\D^0} )
,\end{eqnarray}
which the results of~\cite{Geng:2008mf,Geng:2009hh,Geng:2009ys}
indeed satisfy. 
Results from~\cite{Geng:2009ys} for the magnetic moments of the
$\S^*$, 
moreover, 
are purely isovector. 
This suggests that the low-energy constant
$\ol \mu_{0,\S^*} = 0$, 
which is also what one obtains from matching to 
$SU(3)$ at one-loop order.

\begin{figure}
\begin{tabular}{cc}
\includegraphics[width=0.46\textwidth]{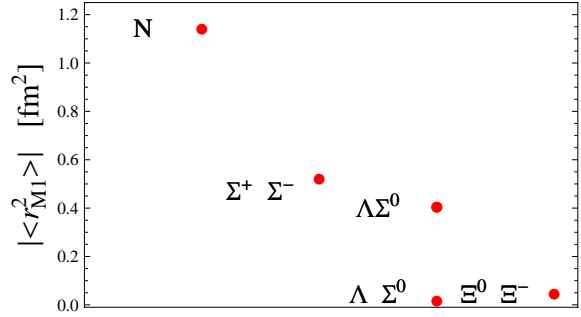}
\\
\end{tabular}
\caption{$SU(2)$ HB\CPT\ results for the octet baryon magnetic radii. 
Note that with Eq.~\eqref{eq:spin1/2}, we do not employ the customary 
definitions for these radii. 
}
\label{f:to_be_made}
\end{figure}

At NLO, 
the magnetic radii for each of the baryons, 
as well as the quadrupole radii of the spin three-half baryons depend on only reasonably known low-energy constants. 
Using these values, 
the 
$SU(2)$
predictions for magnetic radii of the octet baryons are shown in Figure~\ref{f:to_be_made}. 
For the spin three-half resonances, 
however, 
values of the chiral corrections at the pion production threshold are infinite.
Additional physics stemming from pion radiation is required in order to make predictions for physical amplitudes near threshold
(and likely at the physical pion mass too). 
We believe that the singularities encountered in electromagnetic radii and quadrupole moments
deny the resonances these properties,
i.e.~one cannot define such contributions to current matrix elements without also considering pion radiation.
This situation is unlike the case of magnetic moments, 
where the amplitude develops a finite imaginary part associated with the resonance decay.

\begin{figure}
\begin{tabular}{cc}
\includegraphics[width=0.46\textwidth]{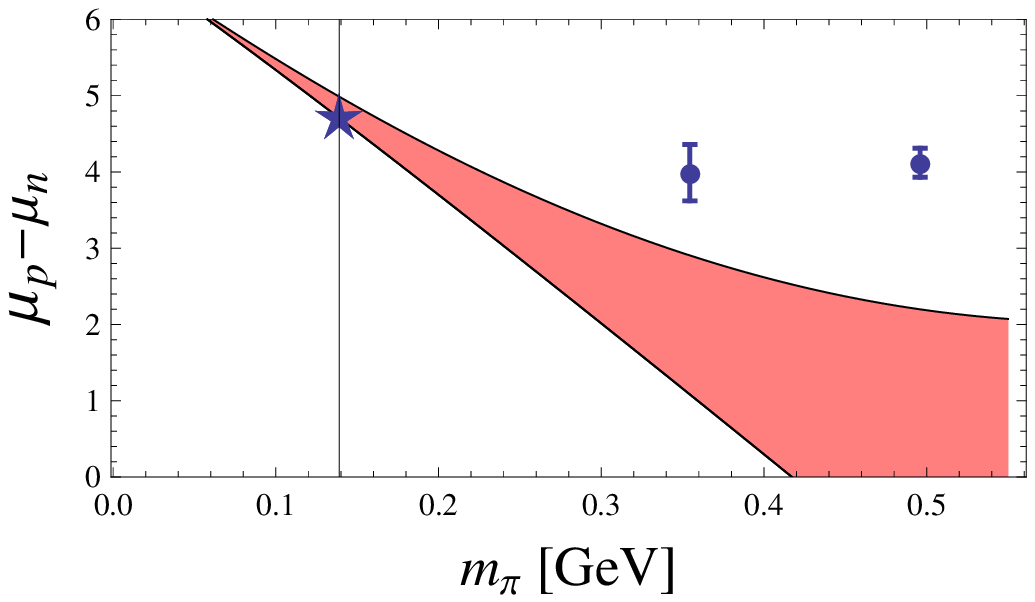}
&\includegraphics[width=0.46\textwidth]{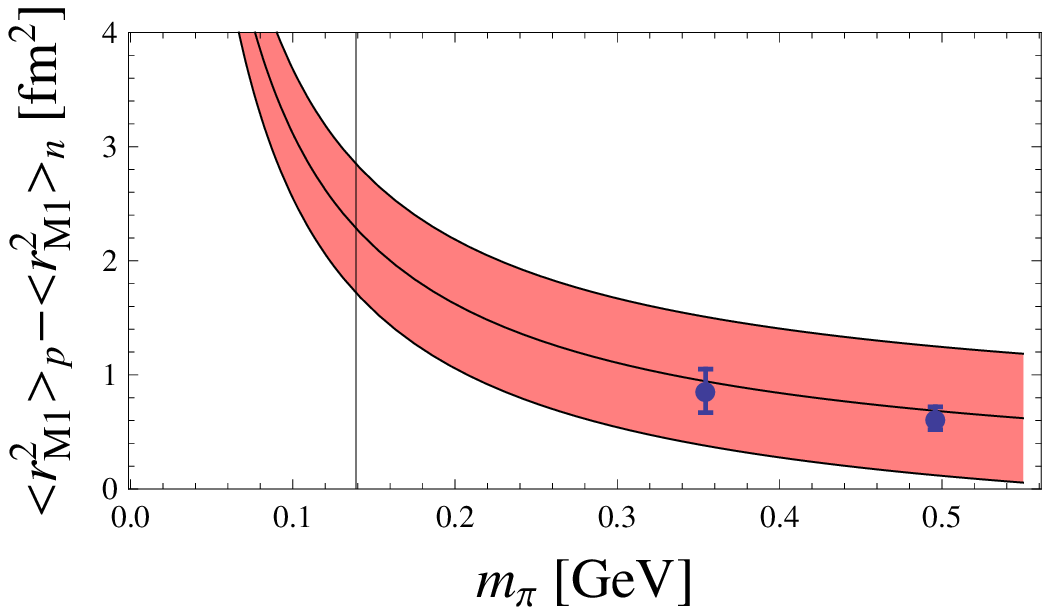}
\\
$(a)$ & $(b)$\\
\includegraphics[width=0.46\textwidth]{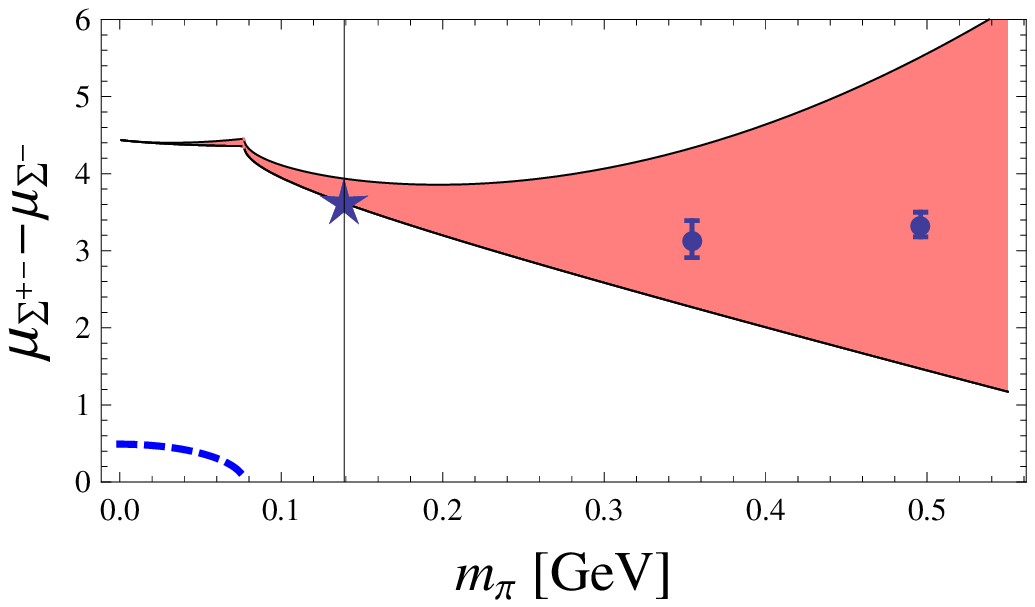}
&\includegraphics[width=0.46\textwidth]{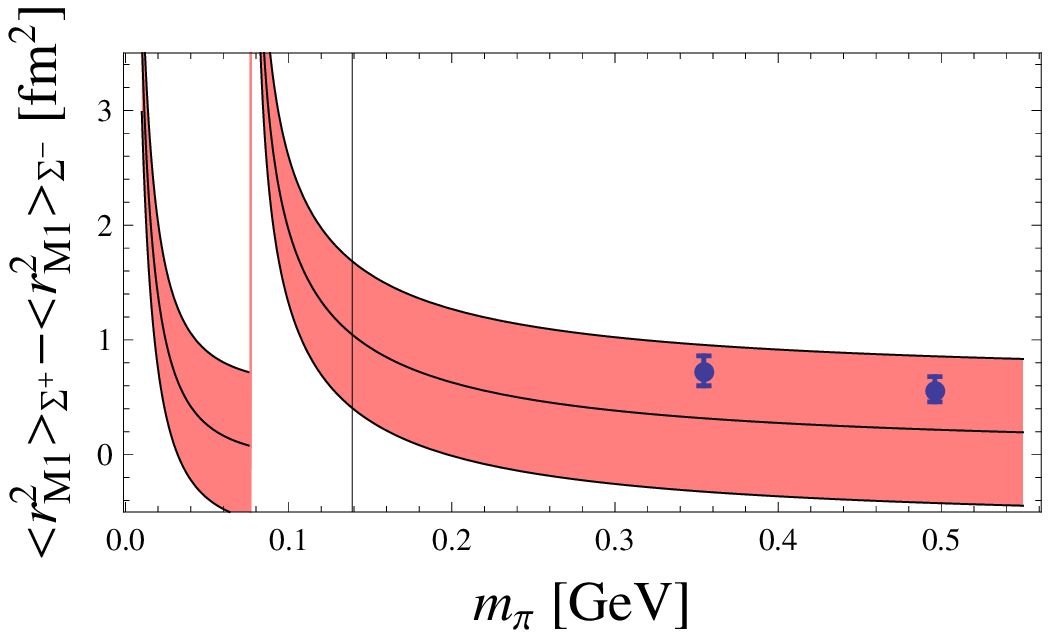}
\\
$(c)$ & $(d)$ \\
\includegraphics[width=0.46\textwidth]{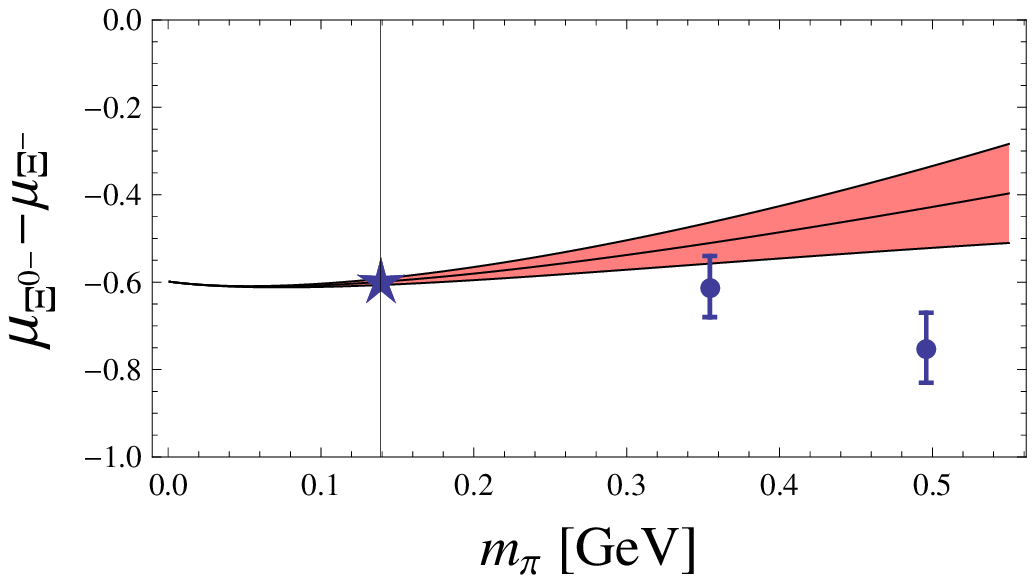}
&\includegraphics[width=0.46\textwidth]{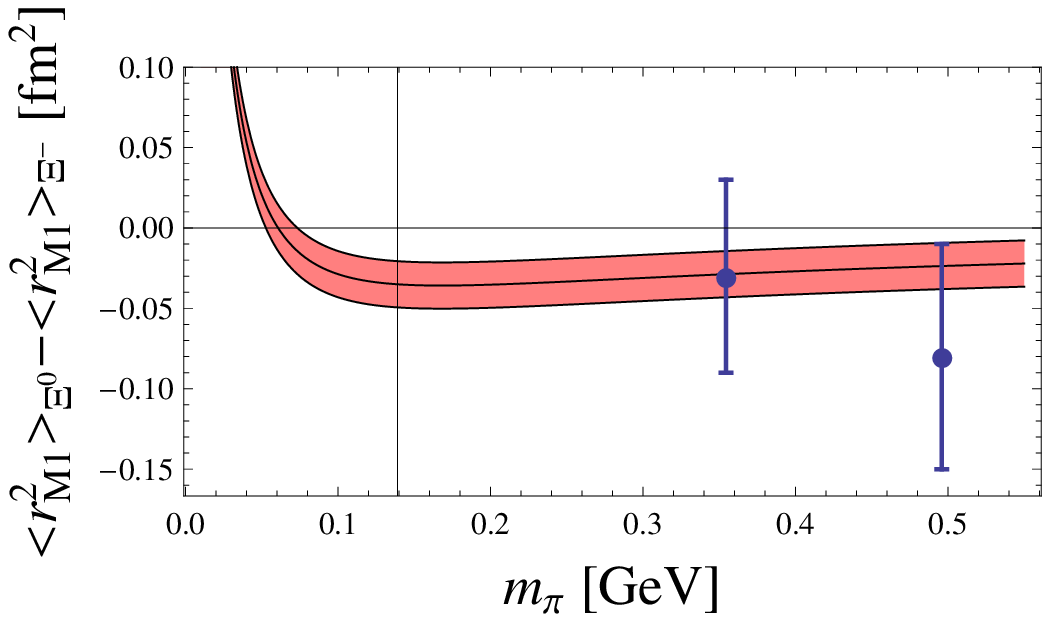}
\\
$(e)$ & $(f)$\\
\includegraphics[width=0.46\textwidth]{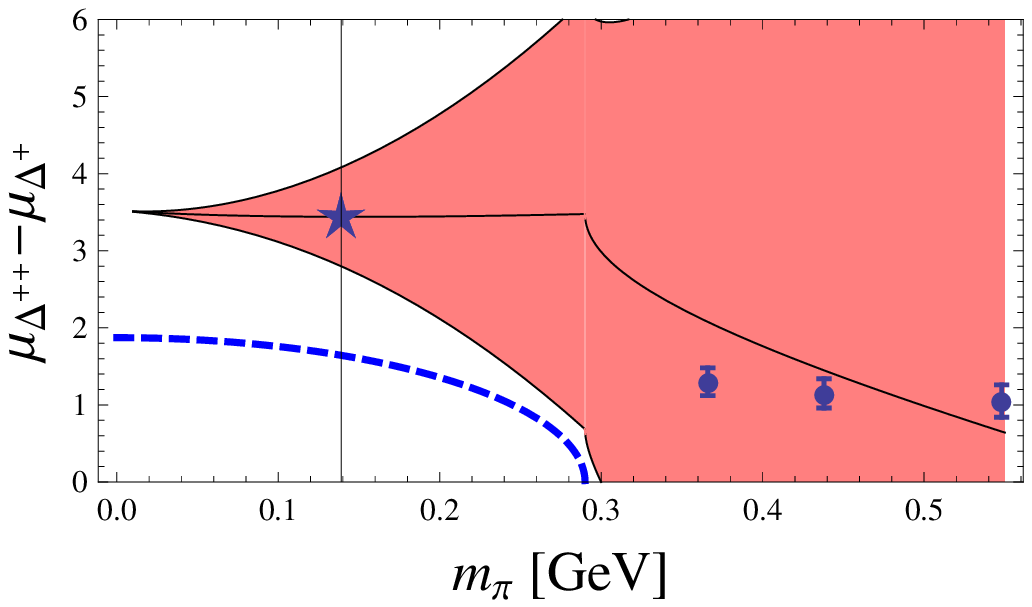}
\\
$(g)$\\
\end{tabular}
\caption{Pion mass dependence of magnetic moments and radii in $SU(2)$. 
The stars represent physical values, 
while the solid circles with error bars are lattice results
of~\cite{Lin:2008mr,Aubin:2008qp}.  
Dashed lines show the imaginary part. 
Uncertainty bands arise from NNLO terms as explained in the text. 
For each observable, the plot range spans the same magnitude 
with the exception of the cascade: 
its magnetic moment plot spans  
$1/6$ 
the range of the other magnetic moment plots, 
while its magnetic radius plot spans 
$1/15$ 
the range of the other magnetic radii plots. 
}
\label{chiralmagnetic}
\end{figure}

We imagine our results to be most useful in comparing with lattice QCD simulations of hyperon properties. 
The expressions we derived above parametrize the pion-mass dependence of the various hyperon electromagnetic properties. 
In principle, we could compare the pion mass dependence of lattice QCD data with that predicted by our formulae. 
This is complicated in practice due to lattice approximations and lattice artifacts. 
In the computation of electromagnetic properties, 
the self-contractions of the current operators are notoriously difficult to calculate due to statistical noise. 
These contributions have been omitted from virtually all computations of current matrix elements.
Fortunately in the strong isospin limit, 
the disconnected parts cancel in 
differences of current matrix elements within an isospin multiplet of fixed strangeness~\cite{Tiburzi:2009yd}.  
Given the available lattice data~\cite{Lin:2008mr,Aubin:2008qp}, 
we can compare our predictions with the magnetic moment differences calculated on the lattice:
$\mu_p - \mu_n$, 
$\mu_{\S^{+}}-\mu_{\S^{-}}$,
$\mu_{\X^{0}}-\mu_{\X^{-}}$, 
and
$\mu_{\D^{++}} - \mu_{\D^{+}}$.
Additionally we can compare our predictions with differences of magnetic radii:%
\footnote{
With Eq.~\eqref{eq:spin1/2}, 
we define the radii to be (six times) the slope of the form factors at 
zero momentum transfer, 
so that differences of radii are independent of sea quark charges in the isospin limit. 
} 
$<r_{M1}^2>_p - <r_{M1}^2>_n$, 
$< r_{M1}^2 >_{\S^{+}}- < r_{M1}^2>_{\S^{-}}$,
and
$< r_{M1}^2>_{\X^{0}}- <r_{M1}^2>_{\X^{-}}$. 
Such comparisons are made in Figure~\ref{chiralmagnetic}.
A final caveat must be issued about the lattice data for spin one-half baryons obtained in~\cite{Lin:2008mr}. 
We have plotted the data at values corresponding to the valence pion mass employed in the simulation. 
The lattice study employs a mixed-action formulation with differing quark actions for the valence and sea quarks. 
Consequently the mixed mesons (consisting of a quark and antiquark from the different fermion discretizations)
are not protected from additive mass renormalization proportional to the lattice spacing squared.
The size of this mass shift has been numerically determined~\cite{Orginos:2007tw,Aubin:2008wk}.
Rather than formulate and perform computations using mixed action \CPT~\cite{Bar:2005tu,Tiburzi:2005is,Chen:2007ug,Chen:2009su} 
to compare with the lattice data, 
we have neglected these discretization effects on the magnetic moments.

In the figure, 
we have included an uncertainty band for our chiral computation. 
Because we have a consistent power counting, 
our computation comes with error estimates from the omitted higher-order terms. 
To obtain error estimates for magnetic moments, 
we have included the analytic term from the NNLO computation.
For magnetic moment differences, 
we have
\begin{eqnarray} 
\d \mu^{NNLO}_{p - n}
=
A_{N}  \frac{8 g_{A}^2 m^2_\pi}{(4 \pi f)^2}, 
&\qquad&
\d \mu^{NNLO}_{\S^+ - \S^-}
=
A_{\S} \frac{4 M_N}{M_\L} \frac{2 g_{\L\S}^2 m_\pi^2}{(4 \pi f)^2},
\notag \\
\label{eq:AB}
\d \mu^{NNLO}_{\X^0 - \X^-}
=
A_{\X} \frac{4 M_N}{M_\X} \frac{2 g_{\X\X}^2 m_\pi^2}{( 4 \pi f)^2}, 
&\qquad&
\d \mu_{\D^{++} - \D^+}^{NNLO} 
= 
A_\D \frac{4 M_N}{M_\D} \frac{2 g_{\D\D}^2 m_\pi^2}{(4 \pi f)^2}
.\end{eqnarray}
A way to estimate the unknown parameters, 
$A_B$,
is to use the fourth-order
$SU(3)$ 
computation of%
~\cite{Meissner:1997hn}. 
The values 
$\d_{p-n}^{NNLO} = 0.11$, 
and 
$\d_{\S^+ - \S^-}^{NNLO} = 0.16$
yield parameters 
$A_N = 1.2$ 
and 
$A_\S = 1.6$
that are of natural size.
Reasonable uncertainty bands are generated by varying
$A_N$ and $A_\S$  in the range $[ 0, 3]$, 
i.e.~letting the 
$SU(3)$ 
result 
vary generously 
$\sim \pm 100\%$.  
The NNLO result for the cascade, 
$\d \mu_{\X^0 - \X^-}^{NNLO} = 0.17$, 
leads to a parameter 
$A_\X$ 
two orders of magnitude greater than natural size. 
This is because the 
$SU(2)$ 
expansion should behave considerably better than the
$SU(3)$
expansion used in%
~\cite{Meissner:1997hn}.
To take this improvement into account, 
we scale the fourth-order
$SU(3)$ 
computation by 
$g_{\X\X}^2 M_N / M_\X$, 
which yields
$A_{\X} = 3.0$. 
As this is an order of magnitude estimate, 
we vary 
$A_{\X}$ 
in the range 
$[-3,3]$. 
As there is no fourth-order computation available for the 
$\D$, 
we shall assume that reasonable variation of 
$A_{\D}$ 
is also in the range
$[-3, 3]$.  
To obtain error estimates for magnetic radii, 
we have included the NNLO analytic terms
\begin{eqnarray} 
\d < r_{M1}^2 >_{p-n}^{NNLO}
&=&
A'_N  
\frac{2 g_{A}^2}{(4 \pi f)^2},
\qquad
\d < r_{M1}^2 >_{\S^+ - \S^-}^{NNLO}
=
A'_\S  
\frac{4 M_N}{M_\L}
\frac{2 g_{\L\S}^2}{(4 \pi f)^2},  
\notag \\
\label{eq:ABB}
\d < r_{M1}^2 >_{\X^0 - \X^-}^{NNLO}
&=&
A'_\X
\frac{4 M_N}{M_\X}
\frac{2 g_{\X \X}^2}{(4 \pi f)^2}
.\end{eqnarray}
The unknown parameters can be estimated by comparing with the fourth-order 
$SU(3)$ 
computation of%
~\cite{Kubis:2000aa}.
At the physical pion mass our computation for the nucleon and sigma are within 
$\sim 15 \%$
of the three-flavor results. 
Inflating this difference by a factor of two as a measure of our uncertainty, 
we find 
$[ -3, 3 ]$ 
constitutes a reasonable range for 
$A'_N$ 
and 
$A'_\S$.
Our value for the isovector magnetic radius of the cascade differs dramatically 
from the value found using 
$SU(3)$. 
This is an observable for which the 
$SU(2)$ 
and 
$SU(3)$
predictions can be tested. 
Assuming naturalness, 
we guess the range
$[-3,3]$
over which to vary $A'_\X$.

The curves in Fig.~\ref{chiralmagnetic} show reasonable agreement with the lattice data. 
Results are generally better for maximal strangeness. 
This is expected as the effectiveness of the effective theory also increases with strangeness. 
For delta observables, 
the large value of the delta axial coupling, 
$g_{\D \D}$, 
may hinder the convergence of 
$SU(2)$.
Higher-order \CPT\ corrections may be needed to address the lattice extrapolation.
To this end, refined values for the hyperon axial couplings are needed, 
e.g.~lattice results for sigma magnetic moments might suggest that the axial couplings in the 
$S=1$ sector are overestimated. 
At this stage, 
it is difficult to provide a definitive reason for the discrepancies. 
On top of effects from the finite lattice volume and lattice spacing,
there are also additional uncertainties in the lattice calculation, 
such as: 
modeling the momentum transfer dependence of form factors, 
or analogously fitting the magnetic field dependence of energies.

\begin{figure}
\begin{tabular}{cc}
 \includegraphics[width=0.46\textwidth]{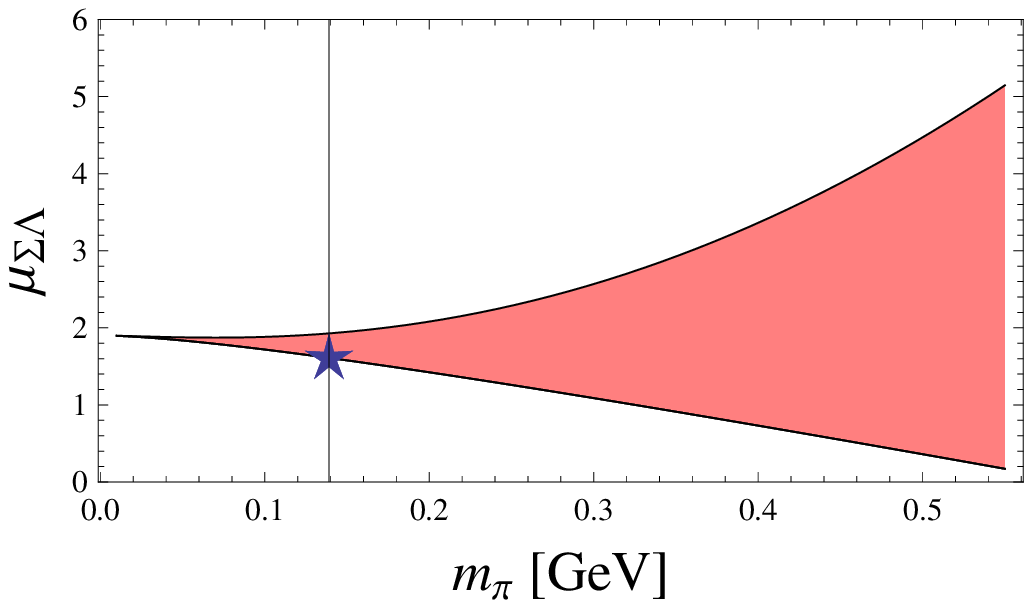}
&\includegraphics[width=0.46\textwidth]{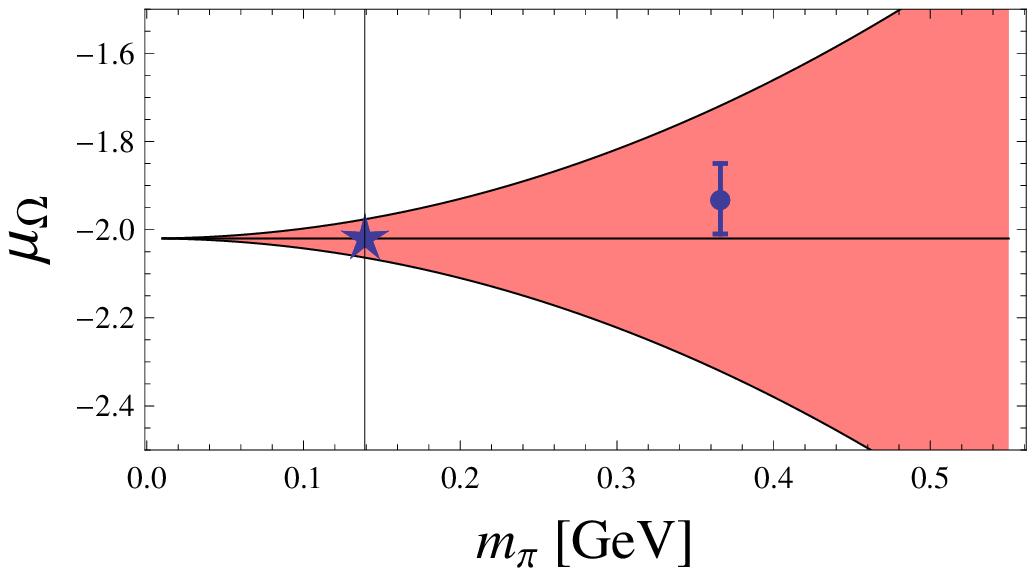}
\\
$(a)$ & $(b)$\\
\includegraphics[width=0.46\textwidth]{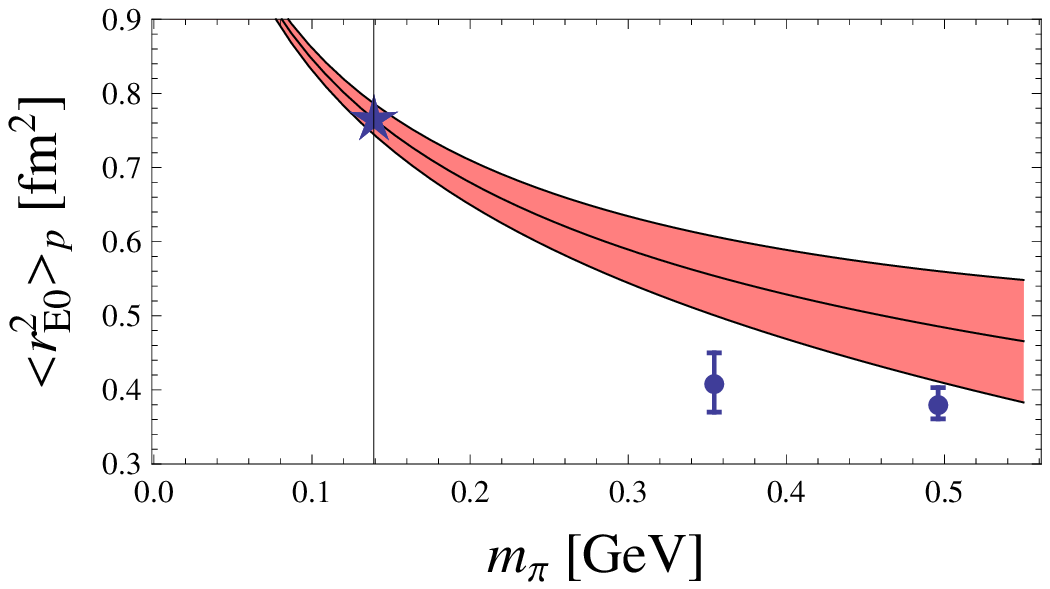}
&\includegraphics[width=0.46\textwidth]{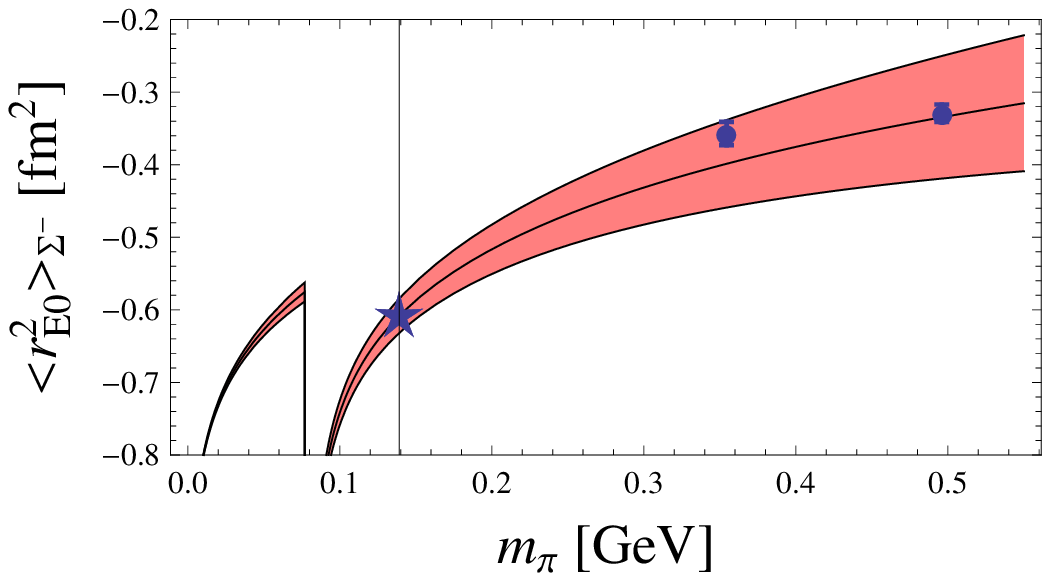}
\\
$(c)$ & $(d)$
\end{tabular}
\caption{
Comparison between 
$SU(2)$ 
HB\CPT\ predictions for the pion mass dependence of baryon electromagnetic properties
and lattice data for their connected parts. 
The stars represent physical values, 
while the solid circles with error bars are lattice results
for charge radii~\cite{Lin:2008mr}, 
and for the omega moment~\cite{Aubin:2008qp}. 
The uncertainty bands on our calculation arise from neglected NNLO terms as explained in the text.
We include the transition moment between the 
$\L$ 
and 
$\S$ 
baryons as an advertisement. 
This magnetic moment has not been determined on the lattice, 
and has \emph{only} connected contributions.  
}
\label{chiral2}
\end{figure}

There exists further lattice data for which we cannot form isospin differences to compare with our formula, 
namely the magnetic moment of the 
$\Omega$, 
and charge radii of the 
$p$, 
and 
$\S^-$.  
For these cases, 
however,
we can still explore the pion mass dependence in light of the data. 
Connected \CPT\ can be employed to determine the modification to pion loop diagrams due
to electrically neutral sea quarks~\cite{Tiburzi:2009yd}, 
however, 
this requires an extension of our work to partially quenched theories. 
We leave this work to future investigation. 
To compare the remaining lattice data with our formulae, 
we assume that the disconnected contributions are negligible. 
If there are considerable differences between our predicted pion mass dependence
and the lattice data, 
it could indicate that disconnected diagrams are important. 
Plots for the remaining electromagnetic properties are shown in Figure~\ref{chiral2}. 
We also show the pion mass dependence of the $\S\L$ transition moment 
which has not been calculated using lattice QCD (and has only connected contributions).

To arrive at the error bands shown in the figure for charge radii, 
we include the analytic term from NNLO,
and estimate its size based on fourth-order 
$SU(3)$ 
computations. 
The analytic term for  
$\S^-$ 
at NNLO has the form
\begin{equation} \label{eq:CB}
\d < r_{E0}^2 >_{\S^-}^{NNLO}
=
C_\S \, \frac{2 g_{\L \S}^2}{(4 \pi f)^2} \frac{m_\pi}{M_\L}
.\end{equation}
From~\cite{Puglia:2000jy}, 
the fourth-order correction has the numerical value 
$\d < r_{E0}^2 >_{\S^-}^{NNLO} = 0.17 [ \texttt{fm}^2]$, 
which corresponds to 
$C_\S = 0.9$, 
for which a reasonable range is 
$[ -3, 3]$. 
As the nucleon charge radii are used as input for the fourth-order analysis, 
we cannot estimate the size of 
$C_N$. 
As a guess, 
we take 
$C_N$ 
also to vary within the range 
$[ -3, 3 ]$. 
For the 
$\S$-$\L$ 
transition moment, 
the 
$SU(3)$ 
computation yields
$\d_{\S\L}^{NNLO}$ 
of the same size as 
$\d_{\S^+ - \S^-}^{NNLO}$%
~\cite{Meissner:1997hn}.
We thus use the same NNLO term to estimate the uncertainty. 
For the magnetic moment of the 
$\O$ 
baryon, 
only local terms enter at NNLO
and lead to 
\begin{equation} \label{eq:AO}
\d \mu_\O^{NNLO} 
= 
A_\O \frac{2 m_\pi^2}{(4 \pi f)^2}
.\end{equation}
As there is no fourth-order calculation available for the 
$\O$, 
we guess the uncertainty band for 
$\mu_\O$
by varying 
$A_\O$ 
in our customary range $[ -3,3 ]$. 
Results for the omega suggest that disconnected diagrams may not be sizable, 
and that our simple prediction for the pion mass dependence can well 
accommodate the data. 
The same appears to be true of the proton's charge radius, 
and of the negative sigma's charge radius.

\section{Conclusion}
\label{conclusion}

In this work, 
we explore two-flavor \CPT\ for hyperons. 
We include electromagnetism into 
$SU(2)$ 
HB\CPT,
and derive the electromagnetic moments and radii of both 
spin one-half and spin three-half hyperons. 
An important aspect of our investigation is to address effects from the nearness of inelastic thresholds. 
To this end, 
we consider the 
$SU(2)$ 
expansion kaon loop contributions. 
We find that the pion mass dependence of kaon loops is well described in 
$SU(2)$ 
for a majority of the hyperon electromagnetic properties. 
Exceptions encountered are the radii and quadrupole moments of the
hyperon resonances, 
for which our results suggest that two-flavor \CPT\ is effective at the physical pion mass, but not much farther.

Using experimental results for spin one-half baryon magnetic moments, 
we are able to deduce values for the 
$SU(2)$ 
low-energy constants. 
Knowledge of these values allows us to compare the size of loop contributions 
relative to the leading local terms. 
We find an improvement in the convergence of 
$SU(2)$ 
over 
$SU(3)$
for most hyperon electromagnetic observables.
We also compare our predictions for the pion mass dependence 
of these electromagnetic observables with lattice QCD data. 
The trends of the data are reasonably captured by our formulae, 
but not without discrepancies. 
We look forward to future lattice data at lower pion masses, 
and larger volumes.
Improvements in lattice QCD calculations will allow us
to refine the values of axial couplings, 
and other low-energy constants appearing in the 
$SU(2)$ 
theory. 
From these values, 
we will be able to demonstrate the convergence pattern of 
\CPT, 
and make predictions for other observables.

\begin{acknowledgments}
B.C.T.~gratefully acknowledges the hospitality of the Institute for Theoretical Physics, Bern University
during an intermediate stage of this work.
Support for this work was provided by the U.S.~Dept.~of Energy, 
under Grants 
No.~DE-FG02-94ER-40818 (F.-J.J.), 
and
No.~DE-FG02-93ER-40762 (B.C.T.). 
\end{acknowledgments}

\appendix

\section*{Appendix: Results from $SU(3)$ HB\CPT}

In this Appendix, we provide expressions for the Coleman-Glashow relations using the three-flavor chiral expansion.  
These expressions depend on three axial couplings: 
those for the octet baryons, 
$D$ 
and 
$F$, 
as well as that of the transition between octet and decuplet baryons, 
$C$. 
We use the standard $SU(3)$ heavy baryon chiral Lagrangian~\cite{Jenkins:1990jv,Jenkins:1991es}.
At NLO, one has the following expressions for the Coleman-Glashow relations~\cite{Jenkins:1992pi}
\begin{eqnarray}
\mu_{\S^-} - \mu_{\X^-}
&=&
\frac{2 M_N }{ 9 \L_\chi^2}
\left[
6  ( D^2 - 6 D F - 3 F^2)  \d F(0)
-
C^2 
\d F(\D)
\right],
\\
\mu_n - 2 \mu_\L
&=&
- 
\frac{4 M_N }{ 9 \L_\chi^2}
\left[
9  ( D + F)^2  \d F(0)
+
2 C^2 
\d F(\D)
\right],
\\
\mu_n - \mu_{\X^0} 
&=&
- 
\frac{4 M_N }{ 9 \L_\chi^2}
\left[
18  ( D^2 + F^2 )  \d F(0)
+
C^2 
\d F(\D)
\right],
\\
\mu_p - \mu_{\S^+} 
&=&
\frac{2 M_N }{ 9 \L_\chi^2}
\left[
6 ( D^2 + 6 D F - 3 F^2)  \d F(0)
+
5 C^2 
\d F(\D)
\right],
\\ 
\mu_n + \mu_{\S^-} + \mu_p
&=&
-
\frac{2 M_N }{ 9 \L_\chi^2}
\left[
12  ( D^2 + 3 F^2)  \d F(0)
-
C^2 
\d F(\D)
\right],
\\
\sqrt{3} \mu_n + 2 \mu_{\S \L} 
&=&
- 
\frac{4 M_N}{ \sqrt{3} \L_\chi^2}
( 3 D^2 - 2 D F + 3 F^2) \d F(0)
.\end{eqnarray}
Notice there are, of course, no local terms to be accounted for in these combinations of magnetic moments. 
We have employed the abbreviation $\L_\chi = 4 \pi f$, where $f$ is the chiral limit meson decay constant. 
Additionally 
$M_N$ 
is the nucleon mass, and appears for each baryon magnetic moment because the moments are given 
in units of nuclear magnetons.
The non-analytic quark mass dependence enters through 
the $SU(3)$ breaking function
$\d F(\Delta)$,
which is given by
$\d F(\D) = F(m_K, \D) - F(m_\pi, \D)$,
where the function $F(m,\d)$ has been given previously in Eq.~\eqref{eq:Ffunc}, 
and $\d F(0) = \pi ( m_K - m_\pi)$.  
To evaluate the Coleman-Glashow relations, 
we use the values~\cite{Jenkins:1992pi}
$D = 0.61$, $F = 0.4$ and $C = 1.2$, 
and take the charged pion and kaon masses, 
along with 
$\D = 0.29 \, \texttt{GeV}$.


\bibliography{hb.bib}


\end{document}